\pdfoutput=1
\documentclass[11pt,a4paper]{article}
\usepackage{jheppub,color,amsfonts,mathrsfs,dsfont,floatrow}
\usepackage[caption=false]{subfig}
\usepackage{tikz}
\usetikzlibrary{decorations.pathreplacing}
\newcommand{\beq}{\begin{eqnarray}}
\newcommand{\eeq}{\end{eqnarray}}
\newcommand{\non}{\nonumber\\}

\newcommand{\p}{\partial}
\DeclareMathOperator{\U}{U}
\DeclareMathOperator{\SU}{SU}
\DeclareMathOperator{\Og}{O}

\DeclareMathOperator{\diag}{diag}

\DeclareMathOperator{\Tr}{Tr}

\DeclareMathOperator{\sign}{sign}

\DeclareMathOperator{\eom}{eom}

\newcommand{\twonorm}[1]{|\mkern-1mu|#1|\mkern-1mu|}

\renewcommand{\i}{\mathrm{i}}
\renewcommand{\d}{{\mathrm{d}}}
\renewcommand{\L}{{\rm L}}
\newcommand{\R}{{\rm R}}

\newcommand{\calM}{\mathcal{M}}
\newcommand{\calV}{\mathcal{V}}
\newcommand{\figsfolder}{figs/}

\newtheorem{theorem}{Theorem}
\newtheorem{corollary}[theorem]{Corollary}

\newtheorem{lemma}[theorem]{Lemma}

\title{Chiral non-Abelian domain walls in the Ginzburg-Landau theory}

\author{Sven Bjarke Gudnason$^1$,}
\affiliation{$^1$Institute of Contemporary Mathematics, School of
  Mathematics and Statistics, Henan University, Kaifeng, Henan 475004,
  P.~R.~China}
\emailAdd{gudnason(at)henu.edu.cn}
\author{Muneto Nitta$^{2,3}$}
\affiliation{$^2$Department of Physics \& Research and Education Center for Natural Sciences,
Keio University, Hiyoshi 4-1-1, Yokohama, Kanagawa 223-8521, Japan}
\affiliation{$^3$International Institute for Sustainability with Knotted Chiral Meta Matter (WPI-SKCM$^2$),
Hiroshima University, 1-3-2 Kagamiyama, Higashi-Hiroshima, Hiroshima 739-8511, Japan}
\emailAdd{nitta(at)phys-h.keio.ac.jp}

\abstract{
In this paper, we study chiral non-Abelian domain walls in a phase of unconventional vacua of the Ginzburg-Landau model for dense QCD, 
by considering a wider range of parameters space not directly deduced from QCD. The phase is characterized by asymmetric vacuum-expectation values (VEVs), for example with the left scalar field, corresponding to the left quark-quark condensate, having a nonvanishing VEV and the right field having a vanishing one. The domain wall soliton interpolates between this vacuum and another where the left and right scalar fields switch roles. 
We study this formal possibility, but not any mechanism to  generate these vacua non-perturbatively at finite density or finite temperature.
Using a strong-coupling, or sigma-model limit, we are able to reduce the full dynamical complex matrix valued equations of motion to the sine-Gordon, a generalization of the sine-Gordon and a generalization of the double sine-Gordon equations.
In this limit, we prove nonexistence of domain walls in one of the vacua studied here and we find full numerical computations to converge to the sigma-model limit for many cases, with some exceptions that we discuss.
}

\begin{document}
\maketitle

\section{Introduction}

Topological defects and solitons, such as monopoles, vortices, and domain walls (DWs), appear and play crucial roles across a wide spectrum of modern physics -- from quantum field theory \cite{Rajaraman:1987,Manton:2004tk,Vachaspati:2006zz,Weinberg:2012pjx,Shnir:2018yzp}, supersymmetric gauge theories \cite{Tong:2005un,Tong:2008qd,Eto:2006pg,Shifman:2007ce,Shifman:2009zz}, and quantum chromodynamics (QCD) \cite{Eto:2013hoa}, to cosmology \cite{Kibble:1976sj,Kibble:1980mv,Vilenkin:1984ib,Hindmarsh:1994re,Vachaspati:2015cma,Vilenkin:2000jqa}, and even in a variety of condensed matter systems \cite{Mermin:1979zz,Volovik:2003fe}.
Among these, DWs (or kinks) represent the simplest type of topological defect and are remarkably ubiquitous in nature \cite{Vachaspati:2006zz,Shnir:2018yzp}. 
In condensed matter physics, for instance, DWs appear in various contexts, such as magnetic DWs in chiral magnets \cite{togawa2012chiral,togawa2016symmetry,KISHINE20151,PhysRevB.97.184303,PhysRevB.65.064433,Ross:2020orc,Amari:2023gqv,Amari:2023bmx,Amari:2024jxx}.
In cosmology, DWs give rise to the so-called DW problem: if they are produced in the early Universe, they can lead to its catastrophic collapse \cite{Kibble:1976sj,Kibble:1980mv,Vilenkin:1984ib,Hindmarsh:1994re,Vachaspati:2015cma,Vilenkin:2000jqa}. 
In quantum field theory, DWs have garnered significant attention from both theoretical and phenomenological perspectives. On the theoretical side, they have been studied extensively in 
supersymmetric nonlinear sigma models 
\cite{Abraham:1992vb,Abraham:1992qv,Gauntlett:2000ib,Tong:2002hi}, 
supersymmetric Yang-Mills theory and in QCD \cite{Dvali:1996xe,Kovner:1997ca,Witten:1997ep,Acharya:2001dz,Isozumi:2004jc,Isozumi:2004va,Eto:2006mz,Eto:2004vy,Bashmakov:2018ghn,
Tong:2005un,Eto:2006pg,Shifman:2007ce,Shifman:2009zz}. 
On the phenomenological side, DWs are of interest in models beyond the Standard Model, including two-Higgs doublet models \cite{Battye:2010dk,Battye:2011jj,Eto:2018hhg,Eto:2018tnk,Chen:2020soj,Battye:2020sxy,Battye:2020jeu,Law:2021ing,Battye:2025nqn}.
Among the many instances, DWs and vortices in QCD have been particularly well studied \cite{Forbes:2000et}. In the presence of chiral symmetry breaking, a rich variety of topological structures emerge, such as axial strings \cite{Balachandran:2001qn,Balachandran:2002je,Nitta:2007dp,Nakano:2007dq,Eto:2009wu,Eto:2013bxa}, axial DWs \cite{Eto:2013bxa,Fukushima:2018ohd}, pion and $\eta$ strings \cite{Zhang:1997is}, pion DWs \cite{Son:2007ny,Eto:2012qd,Brauner:2016pko,Eto:2023lyo,Eto:2023wul,Amari:2024mip,Amari:2024fbo}, and $\eta$ or $\eta'$ DWs \cite{Nishimura:2020odq,Eto:2021gyy,Eto:2023tuu,Eto:2023rzd,Qiu:2023guy}.

In this paper, we focus on DWs in the Ginzburg-Landau theory for quark matter in high-density QCD 
by treating the Ginzburg-Landau parameters as free parameters and considering a parameter region that is not derived from QCD.
In QCD matter, axial $\U(1)_{\rm A}$ DWs have been explored in the two-flavor color superconducting (2SC) phase \cite{Son:2000fh,Buckley:2001bm}, and $\U(1)_{\rm Y}$ DWs in kaon-condensed phases \cite{Son:2001xd,Buckley:2002mx}, see also the review \cite{Eto:2013bxa}.
At asymptotically high densities and low temperatures, QCD is expected to exhibit color superconductivity due to diquark condensation \cite{Alford:2007xm}. 
In this regime, a color-flavor symmetric phase known as the color-flavor locked (CFL) phase arises \cite{Alford:1998mk}, characterized by both superfluid and superconducting properties. The CFL phase involves two diquark condensates -- of left- and right-handed quarks $q_{\rm L,R}$ -- given by
$$(\Phi_{\L,\R})_{\alpha a} \sim \epsilon_{\alpha \beta \gamma} \epsilon_{abc} q_{\L,\R}^{\beta b}  q_{\L,\R}^{\gamma c},$$
where $\alpha,\beta,\gamma=r,g,b$ are color indices, and $a,b,c=u,d,s$ are flavor indices.
In the conventional ground state, both condensates develop vacuum expectation values (VEVs) such that $\Phi_\L = - \Phi_\R$. This configuration supports a variety of topological solitons, including non-Abelian vortices \cite{Eto:2013hoa,Balachandran:2005ev,Nakano:2007dr,Eto:2009kg,Eto:2009bh,Eto:2009tr}, monopoles confined by vortices \cite{Gorsky:2011hd,Eto:2011mk}, and vortex-hadron phase continuity via junctions (boojums) \cite{Cipriani:2012hr,Chatterjee:2018nxe,Chatterjee:2019tbz,Alford:2018mqj}, as well as Higgs–confinement continuity in the presence of vortices \cite{Cherman:2020hbe,Hayashi:2023sas,Cherman:2024exo,Hayata:2024nrl}. 
All of these studies assumed that the condition $\Phi_\L = -\Phi_\R$ holds everywhere, including in the vortex cores. However, vortices without this assumption have recently been explored \cite{Eto:2021nle,Gudnason:2025qxf}, leading to the appearance of non-Abelian sine-Gordon kinks associated with the phase difference between $\Phi_\L$ and $\Phi_\R$ \cite{Nitta:2014rxa}.

In this work, we go a step further by relaxing the assumption of the conventional ground state, and consider the possibility of an unconventional ground state, where either $\Phi_\L$ or $\Phi_\R$ develops a VEV while the other remains uncondensed. 
This scenario emerges naturally within the Ginzburg-Landau (GL) framework with freely tunable parameters, although it lies outside the scope of perturbative QCD analyses valid at asymptotically high densities \cite{Iida:2000ha,Iida:2001pg,Giannakis:2001wz}. 
Such parameter regions can only be generated nonperturbatively, if it exists. 
To better understand this situation, it is useful to draw an analogy with condensed matter systems possessing two scalar condensates, $\Phi_1$ and $\Phi_2$, such as two-component Bose-Einstein condensates (BECs) or two-gap superconductors. 
When both condensates coexist, the system is said to be in a miscible phase. In this case, the overall phase corresponds to an exact $\U(1)$ symmetry, while the relative phase is a global symmetry explicitly broken by a Josephson coupling term, $\Phi_1^* \Phi_2 + \mathrm{c.c.}$
\footnote{In chiral $p$-wave superconductors, the Josephson term takes the form $\Phi_1^{*2} \Phi_2^2 + \mathrm{c.c.}$ \cite{PhysRevLett.80.5184,PhysRevB.86.060514}, giving rise to vortex molecules connected by two DWs.}
This phase difference supports sine-Gordon solitons \cite{Son:2001td,doi:10.1143/JPSJ.70.2844,PhysRevLett.88.017002}, and vortices can form molecular structures consisting of fractional vortices connected by sine-Gordon solitons \cite{Son:2001td,Kasamatsu:2004tvg,Cipriani:2013nya,Tylutki:2016mgy,Eto:2017rfr,Eto:2019uhe,Kobayashi:2018ezm}.
On the other hand, if only one of $\Phi_1$ or $\Phi_2$ condenses, the system enters a phase-separated (immiscible) regime, where two energetically degenerate ground states exist, and DWs can form between them \cite{Ao:1998zz,Timmermans:1997etb,Kasamatsu:2010aq,Nitta:2012hy,Takeuchi:2012ee,Kasamatsu:2013lda,Kasamatsu:2013qia}.
Analogously, in the unconventional CFL ground states of QCD, only one of $\Phi_\L$ or $\Phi_\R$ acquires a VEV. This provides a non-Abelian generalization of the phase-separated structure seen in the aforementioned condensed matter systems.

In this paper, we investigate 
a \emph{chiral non-Abelian} DW interpolating between the two ground states with 
 only one of 
$\Phi_{\rm L}$ and $\Phi_{\rm R}$ developing a VEV. 
Similar non-Abelian DWs were studied in a $\U(N)$ gauge theory
\cite{Shifman:2003uh,Eto:2005cc,Eto:2008dm,Nitta:2015mma,Nitta:2015mxa,Nitta:2022ahj}, but in contrast to those studies, we consider here the Ginzburg-Landau model for dense QCD, which differs both by having two scalar fields (left and right) and by being gauged only in the $\SU(N)$ ($\SU(3)$) part of the left symmetry transformations, as opposed to the entire $\U(N)$ symmetry.
Here, the term ``chiral'' stems from chiral DWs in chiral $P$-wave superconductors \cite{10.1143/PTP.102.965,Bouhon_2010}.
The DW vacua that are the basis of our work here appear if the mixed left-right scalar ``$\phi^4$'' term ($\Tr\Phi_\L\Phi_\L^\dag\Phi_\R\Phi_\R^\dag$) has a larger coefficient, $\lambda_4$, than the pure left or right terms, $\lambda_1$.
We study a formal limit of the Ginzburg-Landau model, where we send both couplings to infinity, but keep the difference fixed and positive: $\lambda_1\to\infty$ with $\lambda_4-\lambda_1=\mathrm{const}>0$.
This sigma-model limit works well for this case, where chiral symmetry is unbroken and in some of the cases where the chiral symmetry is broken.
One term that breaks chiral symmetry to its diagonal is the so-called Josephson term, $\gamma_1\Tr\Phi_\L^\dag\Phi_\R+{\rm c.c.}$
In this paper, we include also the single-trace squared, $\gamma_2\Tr\Phi_\L^\dag\Phi_\R\Phi_\L^\dag\Phi_\R+{\rm c.c.}$ as well as a determinant term $\gamma_3\det\Phi_\L^\dag\Phi_\R+{\rm c.c.}$
Our first results provide more detail on the DW vacua of the Ginzburg-Landau model, of which we have given a preview of in ref.~\cite{Gudnason:2025qxf}, although we restrict ourselves here to the two cases: $\gamma_1\neq0$, $\gamma_2\in\mathbb{R}$ (including zero), $\gamma_3=0$ and $\gamma_1=0$, $\gamma_2\leq0$, $\gamma_3\neq0$.

The second case, where the Josephson ($\gamma_1$) term is turned off is the simplest and is the one where the sigma-model limit yields very good approximations for the true DW solutions, in the limit of relatively large $\lambda_1$.
The first case, with $\gamma_1\neq0$ induces asymmetric ground states, but with both left- and right-scalar fields having nonvanishing VEVs.
It is not approximated too well by the sigma-model limit due to an upper bound on the coupling $\gamma_1\lambda_1<c$, with $c$ a constant, which must be obeyed for the vacua to exist. 
Circumventing this bound by sending $\gamma_1$ to zero as $1/\lambda_1$ in the large $\lambda_1$ limit, has the side effect of reaching a near-fixed point of the equations, that prolongs the DW structure and prevents the sigma-model limit from being an accurate approximation.

On the more formal side, our results include a nonexistence theorem of a DW soliton with the Josephson term turned on ($\gamma_1\neq0$), in the sigma-model limit.
We here adopt a nomenclature of calling the soliton that swaps the left and right scalar fields as DWs, e.g.~a soliton interpolating between $(\Phi_\L,\Phi_\R)=(v\mathds{1}_3,0)$ and $(\Phi_\L,\Phi_\R)=(0,v\mathds{1}_3)$.
In contradistinction, we call a soliton that flips the sign of the VEVs a ``kink'' soliton: it interpolates between $(\Phi_\L,\Phi_\R)=(v_\L,v_\R)\mathds{1}_3$ and $(\Phi_\L,\Phi_\R)=(-v_\L,-v_\R)\mathds{1}_3$.
Although our nonexistence theorem does not allow for the DW soliton in the sigma-model limit, the kink soliton does exist.
The sigma-model limit leads to very simple ordinary differential equations (ODEs) for the profile function of the DW and indeed we find a generalization \eqref{eq:eom_gamma_theta_simpl} of the sine-Gordon equation and a generalization \eqref{eq:eom_gamma1_theta_simpl} of the double sine-Gordon equation \cite{Burt:1977ii,PhysRevB.27.474}, both of which we do not have analytic solutions to.
There exists, however, a fine-tuned point of the Josephson coupling, $\gamma_1$, for which the left and right vacua become degenerate: in this case, the generalization of the double sine-Gordon equation reduces to a special case of the double sine-Gordon equation, for which we find an analytic solution. 
Finally, before turning to numerical computations, we prove that diagonal and real-valued scalar fields lead to soliton solutions with the gauge fields switched off (or pure gauge).

This paper is organized as follows.
In sec.~\ref{sec:CFL}, we review the GL model for the CFL phase and formally study the vacua of the theory with the chiral symmetry breaking terms.
In sec.~\ref{sec:DWs}, we study a strong-coupling limit that we denote the sigma-model limit, for which we find simplified and sometimes analytic exact results.
In sec.~\ref{sec:numerical}, we perform full numerical computations of the matrix equations of motion and compare them to the simplified sigma-model limit.
Finally, we conclude with a discussion and outlook of open problems in sec.~\ref{sec:conclusion}.

\section{Color-flavor-locked phase of 3-flavor Ginzburg-Landau model for dense QCD}\label{sec:CFL}

In this section, we review the GL model for the CFL phase to fix our
notation,
as well as review the ground states from ref.~\cite{Gudnason:2025qxf}
and expand details upon these results.

\subsection{Ginzburg-Landau model for the color-flavor locked phase}

The $\SU(3)_{\rm C}$ color symmetry and  $\U(1)_{\rm B}$ baryon
symmetry are exact while the $\SU(3)_\L\times\SU(3)_\R$ chiral and
$\U(1)_{\rm A}$ axial symmetries are approximate.
These symmetries act on the quark condensates as
\begin{align}
 & \Phi_\L \to 
  e^{\i\varphi_\L} g_{\rm C} \Phi_\L V_\L^\dag , \qquad
  \Phi_\R \to 
  e^{\i\varphi_\R} g_{\rm C} \Phi_\R V_\R^\dag \non
  & g_{\rm C} \in \SU(3)_{\rm C}, \quad 
  U_{\L,\R} \in \SU(3)_{\L,\R}, \quad
  e^{\i\varphi_\L+\i\varphi_\R}\in \U(1)_{\rm B} ,\quad
  e^{\i\varphi_\L-\i\varphi_\R}\in \U(1)_{\rm A}. \label{eq:G-on-Phi}
\end{align}
The vector symmetry $\SU(3)_{\L+\R}$ defined by the condition
$V_\L=V_\R$ is a subgroup of the chiral symmetry
$\SU(3)_\L\times\SU(3)_\R$, and the rest of the generators
parametrize Nambu-Goldstone bosons for the chiral symmetry breaking as
the coset space  
$$[\SU(3)_\L\times\SU(3)_\R]/\SU(3)_{\L+\R}\simeq\SU(3).$$

From here on, we study the Ginzburg-Landau theory of scalar fields with the couplings chosen freely. 
Dense QCD corresponds at a fixed density only to one point in this parameter space of couplings, whereas we study the vacua and solitons in a large part of the parameter space that may not necessarily coincide with physics of QCD at any density.

The static Hamiltonian (energy functional) of the GL model is given
by\footnote{The $\lambda_2$ and $\lambda_3$ terms are normalized as in
ref.~\cite{Gudnason:2025qxf} for convenience, but are divided by $N=3$ as compared
with ref.~\cite{Eto:2021nle}.}
\begin{align}
E &= \frac{1}{2g^2}\twonorm{F}^2 
+ \twonorm{\d_A\Phi_L}^2
+ \twonorm{\d_A\Phi_R}^2
+ V,\label{eq:E}\\
V &= -\frac{m^2}{2}\left(\twonorm{\Phi_\L}^2+\twonorm{\Phi_\R}^2\right)
+ \frac{\lambda_1}{4}\left(\twonorm{\Phi_\L^\dag\Phi_\L}^2+\twonorm{\Phi_\R^\dag\Phi_\R}^2\right)
+ \frac{\lambda_2}{12}\left(\twonorm{\Phi_\L,\Phi_\L}^2 + \twonorm{\Phi_\R,\Phi_\R}^2\right)\non
&\phantom{=\ }
+ \frac{\lambda_3}{6}\twonorm{\Phi_\L,\Phi_\R}^2
+ \frac{\lambda_4}{2}\twonorm{\Phi_\L^\dag\Phi_\R}^2
+ \gamma_1\left(\langle\Phi_\L,\Phi_\R\rangle + \langle\Phi_\L^\dag,\Phi_\R^\dag\rangle\right)\label{eq:V}\\
&\phantom{=\ }
+ \gamma_2\left(\langle\Phi_\R^\dag\Phi_\L,\Phi_\L^\dag\Phi_\R\rangle+\langle\Phi_\L^\dag\Phi_\R,\Phi_\R^\dag\Phi_\L\rangle\right)
+ \gamma_3\int_M\star\left(\det(\Phi_\L^\dag\Phi_\R)+\det(\Phi_\R^\dag\Phi_\L)\right).\nonumber
\end{align}
We define the theory on $\mathbb{R}^d$ with $d=1$, $2$, or $3$; the
kink that we will study in this paper only requires $d=1$ (codimension
of the kink), but taking $d=3$ yields the DWs of our $(3+1)$-dimensional world.
The inner product on $M=\mathbb{R}^d$ ($d=1,2,3$) is defined as
\beq
\langle X,Y\rangle := \Tr\int_M X^\dag\wedge\star Y,
\eeq
where $X,Y$ are both $r$-forms as well as 3-by-3 complex matrices and $\star$ is the Hodge star mapping $r$-forms to ($d-r$)-forms with the property $\star\star=(-1)^{r(d-r)}$. 
The norm squared is then defined as
\beq
\twonorm{X}^2 := \langle X,X\rangle,
\eeq
and finally the double-trace norm squared is given by
\beq
\twonorm{X,Y}^2 = \int_M\Tr(X^\dag\wedge\star X)\wedge\star\Tr(Y^\dag\wedge\star Y).
\eeq
The field-strength 2-form for the SU(3)$_{\rm C}$ color gauge field
and the covariant derivative 1-form are defined as
\begin{align}
  F &:= \d A - A\wedge A = \frac12F_{ij}\d x^{ij},\\
  \d_A\Phi_{\L,\R} &:= d\Phi_{\L,\R} - A \Phi_{\L,\R},
\end{align}
with the notation $\d x^{ij}:=\d x^i\wedge\d x^j$, $i,j=1,2,\ldots,d$ 
and $A=A_i\d x^i$ is an anti-Hermitian 1-form, i.e.~$A^\dag=-A$ and is
$\mathfrak{su}(3)$-valued, which implies that it is also traceless.
If we take $d=1$, the field strength on Euclidean space (spatial
space as opposed to $1+1$ dimensional Minkowski spacetime) vanishes.

For convenience, we provide a few expressions in component form
\begingroup
\allowdisplaybreaks
\begin{align}
\frac{1}{2g^2}\twonorm{F}^2&=-\frac{1}{4g^2}\int_M \Tr F_{ij}F^{ij}\;\d^dx\geq0,\\
\twonorm{\d_A\Phi_\L}^2&=\int_M \Tr (\p_i\Phi_\L-A_i\Phi_\L)^\dag (\p^i\Phi_\L-A^i\Phi_\L)\;\d^dx,\\
\twonorm{\Phi_\L}^2&=\int_M\Tr\Phi_\L^\dag\Phi_\L\;\d^dx,\\
\twonorm{\Phi_\L,\Phi_\R}^2&=\int_M\Tr[\Phi_\L^\dag\Phi_\L]\Tr[\Phi_\R^\dag\Phi_\R]\;\d^dx,\\
\langle\Phi_\L,\Phi_\R\rangle&=\int_M\Tr\Phi_\L^\dag\Phi_\R\;\d^dx,
\end{align}
\endgroup
where spatial indices $i,j$ are lowered (raised) by the (inverse) metric $g_{ij}=\delta_{ij}$ ($g^{ij}=\delta^{ij}$).
Note the negative sign in front of $F_{ij}F^{ij}$ is due to the
anti-Hermitian gauge field, i.e.~$F^\dag=-F$. 

The GL parameters in the GL model in eq.~\eqref{eq:V} were
microscopically calculated in the asymptotically high-density region
of QCD (perturbative QCD) in
refs.~\cite{Iida:2000ha,Iida:2001pg,Giannakis:2001wz}.
In this paper, like in ref.~\cite{Gudnason:2025qxf} we take these
parameters as free GL parameters.

\subsection{First variation}

The equations of motion are given in ref.~\cite{Gudnason:2025qxf}, but
we will display them here for convenience.
The equations are obtained by a first variation of the energy
functional. 
Let $(A_\tau,\Phi_{\L,\tau},\Phi_{\R,\tau})$ be smooth variations of
the fields
$(A,\Phi_\L,\Phi_\R)$ for all $\tau$ and denote by
$\alpha=\p_\tau A_\tau|_{\tau=0}$,
$\beta_\L=\p_\tau\Phi_{\L,\tau}|_{\tau=0}$ and
$\beta_\R=\p_\tau\Phi_{\R,\tau}|_{\tau=0}$.
The first variation is
\begin{align}
\frac{\d}{\d\tau}\bigg|_{\tau=0}\!E
&=\langle\beta_\L,\eom_{\Phi_\L}\rangle_{L^2(M)}
+\langle\eom_{\Phi_\L},\beta_\L\rangle_{L^2(M)}
+\langle\beta_\R,\eom_{\Phi_\R}\rangle_{L^2(M)}
+\langle\eom_{\Phi_\R},\beta_\R\rangle_{L^2(M)}\non
&\phantom{=\ }
+\langle\alpha,\eom_A\rangle_{L^2(M)}
\label{eq:first_variation}
\end{align}
where we have dropped two boundary terms (see
ref.~\cite{Gudnason:2025qxf} for their expression); they are assumed
to vanish, since the covariant derivatives and field strength must
approach zero at spatial infinity in order for the system to have a finite energy.
Requiring the first variation to vanish yields the equations of motion
\begingroup
\allowdisplaybreaks
\begin{align}
\eom_{\Phi_\L}&:=\delta_A\d_A\Phi_\L
-\frac{m^2}{2}\Phi_\L
+\frac{\lambda_1}{2}\Phi_\L\Phi_\L^\dag\Phi_\L
+\frac{\lambda_2}{6}\Phi_\L\Tr(\Phi_\L^\dag\Phi_\L)
+\frac{\lambda_3}{6}\Phi_\L\Tr(\Phi_\R^\dag\Phi_\R)\non
&\phantom{=\ }
+\frac{\lambda_4}{2}\Phi_\R\Phi_\R^\dag\Phi_\L
+\gamma_1\Phi_\R
+2\gamma_2\Phi_\R\Phi_\L^\dag\Phi_\R
+\gamma_3\Xi_\L,\label{eq:eom_PhiL}\\
\eom_{\Phi_\R}&:=\delta_A\d_A\Phi_\R
-\frac{m^2}{2}\Phi_\R
+\frac{\lambda_1}{2}\Phi_\R\Phi_\R^\dag\Phi_\R
+\frac{\lambda_2}{6}\Phi_\R\Tr(\Phi_\R^\dag\Phi_\R)
+\frac{\lambda_3}{6}\Phi_\R\Tr(\Phi_\L^\dag\Phi_\L)\non
&\phantom{=\ }
+\frac{\lambda_4}{2}\Phi_\L\Phi_\L^\dag\Phi_\R
+\gamma_1\Phi_\L
+2\gamma_2\Phi_\L\Phi_\R^\dag\Phi_\L
+\gamma_3\Xi_\R,\label{eq:eom_PhiR}\\
\eom_A&:=\frac{1}{g^2}\left[\delta F+\star A\wedge\star F-\star(\star F\wedge A)\right]
-\d_A\Phi_\L\Phi_\L^\dag+\Phi_\L\d_A\Phi_\L^\dag
-\d_A\Phi_\R\Phi_\R^\dag+\Phi_\R\d_A\Phi_\R^\dag,
\label{eq:eom_A}
\end{align}
\endgroup
which are two 0-forms and a 1-form, respectively.
$\delta$ is the coderivative and $\delta_A$ is the gauge covariant coderivative; writing out the $\delta_A\d_A$ operator we obtain
\beq
\delta_A\d_A\Phi_\L&=-\star\d_A\star\d_A\Phi_\L
=-(\p_i-A_i)(\p^i-A^i)\Phi_\L.
\eeq
Finally, the matrices $\Xi_{\L,\R}$ are variations of the
determinant and are given by
\begin{align}
(\Xi_\L)_{\alpha a}&=\frac12\epsilon_{abc}\epsilon_{def}(\Phi_\R)_{\alpha d}(\Phi_\L^\dag\Phi_\R)_{be}(\Phi_\L^\dag\Phi_\R)_{cf},\non
(\Xi_\R)_{\alpha a}&=\frac12\epsilon_{abc}\epsilon_{def}(\Phi_\L)_{\alpha d}(\Phi_\R^\dag\Phi_\L)_{be}(\Phi_\R^\dag\Phi_\L)_{cf}.
\end{align}
We note that the variation of the determinant can also be written in
the form
\begin{align}
  (\Xi_\L)&=\det(\Phi_\L^\dag\Phi_\R)\Phi_\R(\Phi_\L^\dag\Phi_\R)^{-1},\\
  (\Xi_\R)&=\det(\Phi_\R^\dag\Phi_\L)\Phi_\L(\Phi_\R^\dag\Phi_\L)^{-1},
\end{align}
which is useful only if both $\Phi_\L$ and $\Phi_\R$ have no vanishing
eigenvalues everywhere.

The perturbative masses of the system are:
\beq
m_\Phi=\frac{m}{\sqrt{2}}, \qquad
m_A=g\sqrt{2(v_\L^2+v_\R^2)},
\eeq
with the two VEVs $v_\L=\langle\Phi_\L\rangle$ and
$v_\R=\langle\Phi_\R\rangle$ being determined by the ground state
equations (the equations of motion with vanishing derivatives), see the next subsection.

\subsection{Ground states}

In this section, we will review the ground states found in
ref.~\cite{Gudnason:2025qxf}, but also expand upon those results.
The vacua or ground states of the system are crucially important for
the domain walls (DWs), as they interpolate between them.

\subsubsection{Chirally symmetric \texorpdfstring{($\lambda$)}{(lambda)} ground states}\label{sec:lvac}

Let us first recall the chirally symmetric vacua, which are given by
\begin{align}
\Phi_\L=-\Phi_\R=u\mathds{1}_3, \qquad u=\frac{m}{\sqrt{\lambda_{1234}}},\qquad
\calV=-\frac{3m^4}{2(\lambda_{1234})},\label{eq:uvac}
\end{align}
and
\beq
\left\{\begin{array}{l}
\Phi_\L=v\mathds{1}_3\\\Phi_\R=\mathbf{0}_3
\end{array}\right\}
\ \ \textrm{or}\ \ 
\left\{\begin{array}{l}
\Phi_\L=\mathbf{0}_3\\\Phi_\R=v\mathds{1}_3
\end{array}\right\},\qquad 
v=\frac{m}{\sqrt{\lambda_{12}}},\qquad
\calV=-\frac{3m^4}{4(\lambda_{12})},\label{eq:vvac}
\eeq
where we have introduced the short-hand notation
\beq
\lambda_{i_1\cdots i_n}=\sum_{j=1}^n\sign(i_j)\lambda_{|i_j|};
\label{eq:lambda_def}
\eeq
a negative index corresponds to subtracting that coupling
instead of adding it to the sum.
We use the notation that $V=\int_M\star\calV$, that is, $\calV$ is the energy density of the potential \eqref{eq:V}.
The overall phases have been adjusted by using global $\U(1)_{\rm B}$
and $\U(1)_{\rm A}$ rotations, respectively, see eq.~\eqref{eq:G-on-Phi}.
Two immediate examples are $\lambda_{12}=\lambda_1+\lambda_2$ and
$\lambda_{1234}=\lambda_1+\lambda_2+\lambda_3+\lambda_4$.
The condition for the ground state being either the $v$ or the $u$ ground state is
\begin{align}
\lambda_{12} > \lambda_{34}\ &: \qquad \Rightarrow \qquad
\Phi_\L=-\Phi_\R=u\mathds{1}_3,\\
\lambda_{12} < \lambda_{34}\ &: \qquad \Rightarrow \qquad
\left\{\begin{array}{l}
\Phi_\L=v\mathds{1}_3\\\Phi_\R=\mathbf{0}_3
\end{array}\right\}
\quad\textrm{or}\quad
\left\{\begin{array}{l}
\Phi_\L=\mathbf{0}_3\\\Phi_\R=v\mathds{1}_3
\end{array}\right\},
\end{align}
which correspond to ground states (vacua) without and with DWs,
respectively.
The former is the conventional CFL ground state, but bears no DWs.
On the other hand, the latter are two degenerate, but physically
distinguishable states and hence possess a DW that interpolates
between them.
For a discussion of the plausibility of the condition for this ground
state to exist, see the discussion below.

\subsubsection{Josephson-free chirally broken ground states
  \texorpdfstring{($\gamma_1=0$, $\gamma_2\leq0$, $\gamma_3\neq
    0$)}{(gamma1=0, gamma2<=0, gamma3=/=0)}}\label{sec:vac_gamma23}

It will be convenient to start with the case of $\gamma_1=0$, since
these ground states with chiral symmetry broken either by $\gamma_2$
or by $\gamma_3$ (or both), preserve the vacuum structure -- that is,
the two vacua have one vanishing field each, which is important for the
symmetry considerations.
This will be important for the DW, which we will get to below.

With respect to the $\gamma_3$ vacuum given in
ref.~\cite{Gudnason:2025qxf}, we here include the $\gamma_2$ term as
well.
The solution has the same form as in ref.~\cite{Gudnason:2025qxf}, as
long as $\gamma_2\leq0$:\footnote{We note that the sign
'$\sign(\gamma_3)$' of $\Phi_\R$ was missing in
ref.~\cite{Gudnason:2025qxf} and the condition \eqref{eq:gamma3_vac_lambda12_cond}
was not specified. }
\begin{align}
\Phi_\L&=-\sign(\gamma_3)\Phi_\R=r\mathds{1}_3, \qquad
r=\frac12\sqrt{\frac{\xi-\sqrt{\xi^2-8m^2|\gamma_3|}}{|\gamma_3|}}, \non
\calV &=-\frac{\left(\xi-\sqrt{\xi^2-8m^2|\gamma_3|}\right)\left(16m^2|\gamma_3|-\xi\left(\xi-\sqrt{\xi^2-8m^2|\gamma_3|}\right)\right)}{32\gamma_3^2},
\end{align}
provided that
\beq
\lambda_{12} > \frac{24m^4\gamma_3^2}{\left(\xi-\sqrt{\xi^2-8m^2|\gamma_3|}\right)\left(16m^2|\gamma_3|-\xi\left(\xi-\sqrt{\xi^2-8m^2|\gamma_3|}\right)\right)},
\label{eq:gamma3_vac_lambda12_cond}
\eeq
holds and $\gamma_3$ obeys:
\beq
|\gamma_3|<\frac{\xi^2}{8m^2},
\label{eq:gamma3_cond}
\eeq
where we have defined
\beq
  \xi:=\lambda_{1234}+4\gamma_2,\label{eq:xi_def}
\eeq
and $\lambda$ with many indices is defined in eq.~\eqref{eq:lambda_def}.
It is instructive to expand the above quantities in small $\gamma_3$:
\begin{align}
  r &= \frac{m}{\sqrt{\xi}}\left(1 + \frac{m^2|\gamma_3|}{\xi^2} + \mathcal{O}(|\gamma_3|^2)\right)\\
  \calV &= -\frac{3m^4}{2\xi}\left(1 + \frac{4m^2|\gamma_3|}{3\xi^2}
  +\mathcal{O}(\gamma_3^2)\right),\\
  \lambda_{12} &> \frac{\xi}{2}\left(1 - \frac{4m^2|\gamma_3|}{3\xi^2}
  +\mathcal{O}(\gamma_3^2)\right).
\end{align}

If on the other hand, the condition
\eqref{eq:gamma3_vac_lambda12_cond} is not satisfied, the ground state
becomes instead that of the partially unbroken phase: 
\beq
\left\{\begin{array}{l}
\Phi_\L=v\mathds{1}_3\\\Phi_\R=\mathbf{0}_3
\end{array}\right\}
\ \ \textrm{or}\ \ 
\left\{\begin{array}{l}
\Phi_\L=\mathbf{0}_3\\\Phi_\R=v\mathds{1}_3
\end{array}\right\},\qquad 
v=\frac{m}{\sqrt{\lambda_{12}}},\qquad
\calV=-\frac{3m^4}{4\lambda_{12}};
\label{eq:vac_gamma3_dw}
\eeq
this is the ground state that bears DWs.
We should comment on that the $\gamma_3$ term theoretically equips the
potential with runaway directions, that can be triggered for very
large field values.
Physically, however, this must be just an artifact of the low-energy
Effective Field Theory (EFT).

For $\gamma_2>0$, we are not able to solve the vacuum equations
analytically for the exact value of the complex phases of the fields.

\subsubsection{Josephson chirally broken ground states
  \texorpdfstring{($\gamma_1\neq0$, $\gamma_2\in\mathbb{R}$, $\gamma_3=0$)}{(gamma1=/=0, gamma2 in R, gamma3=0)}}

We will now turn on the Josephson term: $\gamma_1\neq0$.
As found in ref.~\cite{Gudnason:2025qxf}, there will still be two
competing ground states that depend on whether
$\lambda_3+\lambda_4$ is large or not.
Due to the linear term (proportional to $\gamma_1$) in the ground
state equations, the vanishing VEV from before is shifted slightly,
turning on a finite VEV for the field that vanished in the absence of
the Josephson term in the DW vacuum. 
We include $\gamma_2$ here for compactness of the presentation, but it
is allowed to vanish.
We have \cite{Gudnason:2025qxf}:
\begin{align}
\Phi_\L&=-\sign(\gamma_1)\Phi_\R=w\mathds{1}_3, \qquad
w=\sqrt{\frac{m^2+2|\gamma_1|}{\xi}},\non
\calV&=-\frac{3(m^2+2|\gamma_1|)^2}{2\xi},
\end{align}
with $\xi$ defined in eq.~\eqref{eq:xi_def}.
This is the ground state (global minimum) when the
condition\footnote{We note that there were some typos in both the
condition \eqref{eq:gamma12cond} as well as in the ground state,
$w_\pm$ of eq.~\eqref{eq:vac_gamma1_dw} in ref.~\cite{Gudnason:2025qxf}.}
\beq
2\left(1+\frac{2|\gamma_1|}{m^2}\right)^2 > \left(1+\frac{\lambda_{34}+4\gamma_2^2}{\lambda_{12}}\right)\left(1+\frac{8\gamma_1^2\lambda_{12}}{m^4(\lambda_{34-1-2}+4\gamma_2)}\right),
\label{eq:gamma12cond}
\eeq
is satisfied and otherwise the following is the ground state
\begin{align}
\Phi_\L&=w_\pm\mathds{1}_3,\qquad
\Phi_\R=-\sign(\gamma_1)w_\mp\mathds{1}_3,\non
w_\pm&=\frac{m}{\sqrt{2\lambda_{12}}}\sqrt{1\pm
\sqrt{1-\frac{16\gamma_1^2\lambda_{12}^2}{m^4(\lambda_{34-1-2}+4\gamma_2)^2}}},\non
\calV&=-\frac{3m^4}{4\lambda_{12}}-\frac{6\gamma_1^2}{\lambda_{34-1-2}+4\gamma_2}.
\label{eq:vac_gamma1_dw}
\end{align}
Notice that when $\gamma_1\to0$, the $w_-$ VEV vanishes while 
$w_+$ reduces to $\frac{m}{\sqrt{\lambda_{12}}}=v$, which is exactly the
Josephson-less DW ground state of eq.~\eqref{eq:vvac} (and also
eq.~\eqref{eq:vac_gamma3_dw}).
We also notice that setting $\gamma_1:=0$, the condition
\eqref{eq:gamma12cond} reduces to
$\lambda_{12}>\lambda_{34}+4\gamma_2$, which for $\gamma_2:=0$ is the
condition without the Josephson term, i.e.~$\lambda_{12}>\lambda_{34}$. 
This asymmetric ground state \eqref{eq:vac_gamma1_dw} is
radically different from the asymmetric ground states \eqref{eq:vvac}
and \eqref{eq:vac_gamma3_dw} by a small shift in the vanishing field
by the linear (in the equation of motion) Josephson term, for
$\gamma_1\neq0$.
Expanding the VEVs in small $\gamma_1$, we have
\begin{align}
w_+ &= \frac{m}{\sqrt{\lambda_{12}}}\left(1 - \frac{2\gamma_1^2\lambda_{12}^2}{m^4(\lambda_{34-1-2}+4\gamma_2)^2} + \mathcal{O}(\gamma_1^4)\right),\\
w_- &= \frac{2|\gamma_1|\sqrt{\lambda_{12}}}{m|\lambda_{34-1-2}+4\gamma_2|} + \mathcal{O}(\gamma_1^2).
\end{align}
In particular, global (flavor) transformations will rotate the field,
which without the presence of the $\gamma_1$ term vanishes in these
ground states.
This reduces the amount of simplification that we may impose on the
model by symmetry considerations, see below.

\section{Domain walls in sigma-model limit}\label{sec:DWs}

Static gauge fields depending only on one dimension in the absence of electric charges leads to a vanishing field strength tensor, $F=0$.
This is easy to see, since the absence of electric sources allows us to set $A_0=0$ and there is no nonvanishing antisymmetric structure in one dimension.
Alternatively, one could consider the strong gauge coupling limit, that would eliminate the field strength tensor from the Lagrangian (and from the energy functional).
In either case, the equation of motion for the gauge field reduces to
\begin{equation}
\{A_j,\Phi_\L\Phi_\L^\dag+\Phi_\R\Phi_\R^\dag\}
-\p_j\Phi_\L\Phi_\L^\dag
+\Phi_\L\p_j\Phi_\L^\dag
-\p_j\Phi_\R\Phi_\R^\dag
+\Phi_\R\p_j\Phi_\R^\dag=0,
\label{eq:eomA_ginfty}
\end{equation}
where it is understood that this equation holds only for the traceless
part, since $A\in\mathfrak{su}(3)$.
This determines the anti-commuting part of $A_j$ implicitly (an explicit expression can be found using the group structure of $\SU(3)$).

It will prove convenient to rewrite the potential in order to define a ``sigma-model limit''.
In particular, we define $H=(\Phi_\L,\Phi_\R)$ and notice that
\begin{align}
\frac{\lambda_1}{4}\Tr\big(H H^\dag-v^2\mathds{1}_{3}\big)^2
&=-\frac{\lambda_1 v^2}{2}\Tr\big(\Phi_\L^\dag\Phi_\L+\Phi_\R^\dag\Phi_\R\big)
+\frac{\lambda_1}{4}\Tr\left[(\Phi_\L^\dag\Phi_\L)^2+(\Phi_\R^\dag\Phi_\R)^2\right]\non
&\phantom{=\ }
+\frac{\lambda_1}{2}\Tr\big(\Phi_\R^\dag\Phi_\L\Phi_\L^\dag\Phi_\R\big)
+\frac{3\lambda_1v^4}{4},
\end{align}
whereas
\begin{align}
\frac{\lambda_2}{12}\big(\Tr[H H^\dag]-3\tilde{v}^2\big)^2
&=-\frac{\lambda_2\tilde{v}^2}{2}\Tr\big(\Phi_\L^\dag\Phi_\L+\Phi_\R^\dag\Phi_\R\big)
+\frac{\lambda_2}{12}\left(\Tr\big(\Phi_\L^\dag\Phi_\L\big)^2+\Tr\big(\Phi_\R^\dag\Phi_\R\big)^2\right)\non
&\phantom{=\ }
+\frac{\lambda_2}{6}\Tr\big(\Phi_\L^\dag\Phi_\L\big)\Tr\big(\Phi_\R^\dag\Phi_\R\big)
+\frac{3\lambda_2\tilde{v}^4}{4}.
\end{align}
We can thus write the $\lambda$-part of the potential as
\begin{align}
\calV &= 
\frac{\lambda_1}{4}\Tr\big(H H^\dag-v^2\mathds{1}_{3}\big)^2
+\frac{\lambda_2}{12}\big(\Tr[H H^\dag]-3\tilde{v}^2\big)^2
+\frac{\lambda_4-\lambda_1}{2}\Tr\big(\Phi_\L^\dag\Phi_\L\Phi_\R^\dag\Phi_\R\big)\non
&\phantom{=\ }
+\frac{\lambda_3-\lambda_2}{6}\Tr\big(\Phi_\L^\dag\Phi_\L\big)\Tr\big(\Phi_\R^\dag\Phi_\R\big)
-\frac{3\lambda_1v^4}{4}
-\frac{3\lambda_2\tilde{v}^4}{4},
\end{align}
where the constants $v$ and $\tilde{v}$ are related to the mass as
\beq
\lambda_1v^2+\lambda_2\tilde{v}^2=m^2.
\label{eq:uvrel}
\eeq
Sending $\lambda_1\to\infty$ imposes the constraint
\beq
H H^\dag = v^2\mathds{1}_3,
\label{eq:unitary_constraint}
\eeq
since the vanishing of the trace of a Hermitian matrix squared implies
that the matrix vanishes. 
Whereas sending $\lambda_2\to\infty$ imposes only the much weaker constraint
\beq
\Tr[H H^\dag]=3\tilde{v}^2,
\eeq
which only fixes the norm, but does not require the components of $H$
to be a unitary matrices\footnote{Strictly speaking, the condition
$\Phi_\L^\dag\Phi_\L+\Phi_\R^\dag\Phi_\R=v^2\mathds{1}_3$ does not
require $\Phi_{\L,\R}$ to be unitary matrices.
However, we assume here that the vacua are such that either $\Phi_\L$
or $\Phi_\R$ vanishes, which thus forces the other field to be a
unitary matrix-valued field. When the fields interpolate between two
such vacua, both fields must thus be unitary matrix-valued fields (see, however, the discussion below). 
}. 
We, therefore, want to send $\lambda_1$ to infinity and use the freedom of the relation between $\tilde{v}$ and $v$ of eq.~\eqref{eq:uvrel} to set $\tilde{v}^2=v^2$.
We thus consider the limit
\beq
\lambda_1\to\infty, \qquad
|\lambda_4-\lambda_1| \qquad {\rm finite}.
\label{eq:sigma_model_limit}
\eeq
This ensures that the potential for the matrix $H$ in the sigma-model limit reduces to
\begin{align}
\calV^{\sigma} &= 
\frac{\lambda_4-\lambda_1}{2}\Tr\big(\Phi_\R^\dag\Phi_\L\Phi_\L^\dag\Phi_\R\big)
+\frac{\lambda_3-\lambda_2}{6}\Tr\big(\Phi_\L^\dag\Phi_\L\big)\Tr\big(\Phi_\R^\dag\Phi_\R\big)
-\frac{3m^2v^2}{4}.
\end{align}
The constraint \eqref{eq:unitary_constraint} implies that
\beq
\Phi_\L\Phi_\L^\dag + \Phi_\R\Phi_\R^\dag = v^2\mathds{1}_3.
\label{eq:sigma_model_constraint}
\eeq
We can thus parametrize the fields as
\beq
\Phi_\L = vU e^{\i\varphi_U}\cos\theta,\qquad
\Phi_\R = vV e^{\i\varphi_V}\sin\theta,
\eeq
with $U$ and $V$ being two special unitary matrices and $\theta$,
$\varphi_U$ and $\varphi_V$ parametrize the unit 3-sphere allowed by
the constraint \eqref{eq:sigma_model_constraint}.
Inserting this parametrization into the potential yields
\beq
\calV^{\sigma}=\frac{3\lambda_{34-1-2}v^4}{8}\sin^2(2\theta),
\eeq
where we have dropped a constant term.
We can also see that this potential is positive definite, when $\lambda_1+\lambda_2<\lambda_3+\lambda_4$, which is indeed the criterion for the DW vacua to appear.

We will now promote the two unitary matrices to fields dependent only
on the $x^1$ direction (and denote the coordinate as $x$):
\beq
\Phi_\L = vU(x) e^{\i\varphi_U(x)}\cos\theta(x),\qquad
\Phi_\R = vV(x) e^{\i\varphi_V(x)}\sin\theta(x),
\label{eq:DWansatz_raw}
\eeq
We contemplate the following situation where this field setup describes a domain wall between two different vacua:
\begin{align}
U(-\infty)&=U^{-\infty},& 
V(\infty)&=V^\infty,\label{eq:BC_pre1}\\
\varphi_U(-\infty)&=\varphi_U^{-\infty},&
\varphi_V(\infty)&=\varphi_V^{\infty}\\
\theta(-\infty)&=0,&
\theta(\infty)&=\frac\pi2,\label{eq:BC_pre3}
\end{align}
which correspond to the fields
\begin{align}
    \Phi_\L(-\infty)&=v U^{-\infty} e^{\i\varphi_U^\infty},&\Phi_\L(\infty)&=0,\non
    \Phi_\R(-\infty)&=0,&\Phi_\R(\infty)&=v V^\infty e^{\i\varphi_V^\infty},
\end{align}
which clearly obey the constraint \eqref{eq:sigma_model_constraint}.

Performing a gauge transformation as well as both a left- and right-handed flavor transformation, we have
\begin{equation}
\Phi_\L = v e^{\i(\varphi_\L+\varphi_U(x))} g_C(x)U(x)V_\L^\dag\cos\theta(x),\qquad
\Phi_\R = v e^{\i(\varphi_\R+\varphi_V(x))} g_C(x)V(x)V_\R^\dag\sin\theta(x).
\end{equation}
We can gauge away the $\SU(3)$-part of $\Phi_R$ using 
\beq
g_C(x) = V_\R V(x)^\dag.
\eeq
For convenience, we have included the constant matrix $V_\R$, so that the right field is proportional to the unit matrix:
\beq
\Phi_\L = v e^{\i(\varphi_\L+\varphi_U(x))} V_\R V(x)^\dag U(x)V_\L^\dag\cos\theta(x),\qquad
\Phi_\R = v e^{\i(\varphi_\R+\varphi_V(x))} \mathds{1}_3\sin\theta(x).
\eeq
Evaluating the fields at the two extremes now yields
\begin{align}
\Phi_\L(-\infty)&=v e^{\i(\varphi_\L+\varphi_U^{-\infty})}V_\R V(-\infty)^\dag U^{-\infty} V_\L^\dag,&
\Phi_\L(\infty)&=0,\non
\Phi_\R(-\infty)&=0,&
\Phi_\R(\infty)&=v e^{\i(\varphi_\R+\varphi_V^\infty)}\mathds{1}_3.
\end{align}
We are free to choose the global flavor rotations to simplify the
boundary conditions as much as possible; in particular, we may choose
$\varphi_\L:=-\varphi_U^{-\infty}$, $V_\L:=V_\R V(-\infty)^\dag U^{-\infty}$,
$\varphi_\R:=-\varphi_V^\infty$.

We may simplify the notation by defining
$W(x):=V_\R V(x)^\dag U(x)V_\L^\dag$
which is a special unitary field,
$\psi(x):=\varphi_\L+\varphi_U(x)$ 
and $\phi(x):=\varphi_\R+\varphi_V(x)$ which are both real valued.
The field Ansatz for the domain wall thus becomes
\beq
\Phi_\L = v e^{\i\psi(x)} W(x) \cos\theta(x),\qquad
\Phi_\R = v e^{\i\phi(x)} \mathds{1}_3\sin\theta(x),
\label{eq:DWansatz}
\eeq
with the boundary conditions
\begin{align}
    W(-\infty) &=\mathds{1}_3, & 
    \phi(\infty) &= 0,\label{eq:BC_Wphi}\\
\theta(-\infty)&=0,&
\theta(\infty)&=\frac\pi2,\label{eq:BC_theta}\\
\psi(-\infty)&=0,\label{eq:BC_psi}
\end{align}
which correspond to the fields
\begin{align}
    \Phi_\L(-\infty)&=v \mathds{1}_3,&\Phi_\L(\infty)&=0,\non
    \Phi_\R(-\infty)&=0,&\Phi_\R(\infty)&=v \mathds{1}_3.
\end{align}
Clearly the sigma-model constraint \eqref{eq:sigma_model_constraint} and the vacuum conditions are obeyed.
We thus have a special unitary matrix $W(x)$, two real-valued
functions $\psi(x)$ and $\phi(x)$, as well as the domain wall profile
$\theta(x)$.

Returning now to the equation of motion for the gauge field, we have
\beq
\left(
\{A_j,\Phi_\L\Phi_\L^\dag+\Phi_\R\Phi_\R^\dag\}
-\p_j\Phi_\L\Phi_\L^\dag
+\Phi_\L\p_j\Phi_\L^\dag
-\p_j\Phi_\R\Phi_\R^\dag
+\Phi_\R\p_j\Phi_\R^\dag\right)\Big|_{\rm traceless}=0,
\label{eq:eomA}
\eeq
where $X|_{\rm traceless}$ is a short-hand notation for the projection
$\tfrac12\Tr[X\lambda^a]\lambda^a$, $a=1,2,\ldots,8$, where $\lambda^a$ are the
Gell-Mann matrices -- the $\SU(3)$ generators. 
It is worth to notice that this equation is invariant under both gauge and flavor transformations.
Inserting the sigma-model constraint \eqref{eq:sigma_model_constraint} and using the unitary property of the fields, we can explicitly determine the gauge field as
\beq
A = \frac{\d\Phi_\L\Phi_\L^\dag+\d\Phi_\R\Phi_\R^\dag}{v^2}\bigg|_{\rm traceless}.
\label{eq:Asol_general}
\eeq
Using now the DW Ansatz \eqref{eq:DWansatz}, we obtain
\beq
A = \d W W^{-1}\cos^2\theta,
\label{eq:Asol_sigma_model_limit}
\eeq
which is traceless due to the special unitary property of $W$.

Computing the kinetic terms (the covariant derivatives of $\Phi_\L$ and $\Phi_\R$) for the theory in the sigma-model limit gives
\begin{align}
  \|\d_A\Phi_\L\|^2+\|\d_A\Phi_\R\|^2&=
  \frac{v^2}{4}\|\d W\sin(2\theta)\|^2
+v^2\|\d\psi\cos\theta\mathds{1}_3\|^2
+v^2\|\d\phi\sin\theta\mathds{1}_3\|^2\non
&\phantom{=\ }
+v^2\|\d\theta\mathds{1}_3\|^2,
\end{align}
where we retain $\mathds{1}_3$ to remember that the trace gives a factor of 3.
The total energy functional in the sigma-model limit thus reads
\begin{align}
E&=\frac{v^2}{4}\|\d W\sin(2\theta)\|^2
+v^2\|\d\psi\cos\theta\mathds{1}_3\|^2
+v^2\|\d\phi\sin\theta\mathds{1}_3\|^2
+v^2\|\d\theta\mathds{1}_3\|^2\non
&\phantom{=\ }
+\frac{\lambda_{34-1-2}v^4}{8}\|\sin(2\theta)\mathds{1}_3\|^2,
\label{eq:Esigmamodel_lambda}
\end{align}
up to an additive constant.
Energy minimization thus gives rise to
\beq
\d W = 0, \qquad
\d\psi = 0, \qquad
\d\phi = 0,
\label{eq:constant_W_psi_phi}
\eeq
as well as the sine-Gordon equation
\beq
\delta\d\theta + \frac{\lambda_{34-1-2}v^2}{8}\sin(4\theta)=0,
\label{eq:SG}
\eeq
although the full equations of motion are more complicated:
\begin{align}
  \delta\left(\d W W^\dag\sin^2(2\theta)\right)&=0,\label{eq:eom_W}\\
  \delta\left(\d\psi\cos^2\theta\right)&=0,\label{eq:eom_psi}\\
  \delta\left(\d\phi\sin^2\theta\right)&=0,\label{eq:eom_phi}\\
  \delta\d\theta
  +\frac12\left(\d\phi\wedge\star\d\phi-\d\psi\wedge\star\d\psi\right)\sin(2\theta)\qquad\non
  \mathop+\frac14\left(\frac{\lambda_{34-1-2}v^2}{2}+\frac13\Tr[\d
    W^\dag\wedge\star\d W]\right)\sin(4\theta)&=0.\label{eq:eom_theta}
\end{align}
Although one could think of contrived solutions where the arguments of
$\delta$ in eqs.~\eqref{eq:eom_W}-\eqref{eq:eom_phi} is a constant
with nonvanishing $\d W$, $\d\psi$ and $\d\phi$; such a solution would
imply infinite derivatives at $\theta=0$ or $\theta=\frac\pi2$.
We thus need to impose eq.~\eqref{eq:constant_W_psi_phi}, for which
the equation of motion for $\theta$ reduces to the sine-Gordon
equation \eqref{eq:SG}.

Now since we have established that $W$, $\psi$ and $\phi$ are
constants, they are indeed determined by the boundary conditions
\eqref{eq:BC_Wphi} and \eqref{eq:BC_psi}.
The condition $W=\mathds{1}_3$ translates to
\beq
V(x)^\dag U(x) = V(-\infty)^\dag U^{-\infty},
\eeq
where we observe that $V_\R$ drops out.
By construction, in the limit $x\to-\infty$ this condition holds true.
But due to the condition that $W$ be constant for all $x$, the
solution is simply $U(x)=V(x)V(-\infty)^\dag U^{-\infty}$; we see that the left- and right-handed
fields are locked together in the DW solution, leaving it with 8 real
'rotational' moduli.

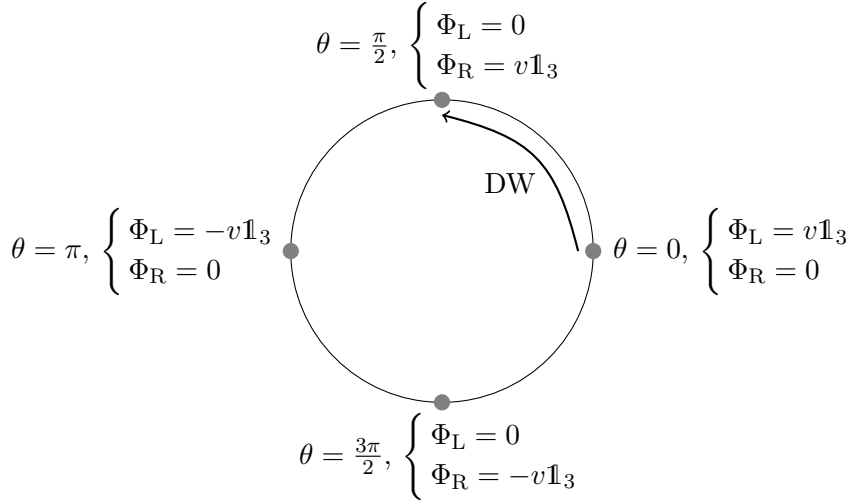
\begin{figure}[!ht]
  \begin{center}
    \begin{tikzpicture}[scale=1.0]
      \draw (0,0) circle (2);
      \filldraw [gray] (2,0) circle (0.1);
      \filldraw [gray] (0,2) circle (0.1);
      \filldraw [gray] (-2,0) circle (0.1);
      \filldraw [gray] (0,-2) circle (0.1);
      \draw (2,0) node [right] {$\ \theta=0$, $\left\{\begin{array}{l}\Phi_\L=v\mathds{1}_3\\\Phi_\R=0\end{array}\right.$};
      \draw (0,2) node [anchor=south] {$\theta=\frac{\pi}{2}$, $\left\{\begin{array}{l}\Phi_\L=0\\\Phi_\R=v\mathds{1}_3\end{array}\right.$};
      \draw (-2,0) node [left] {$\theta=\pi$, $\left\{\begin{array}{l}\Phi_\L=-v\mathds{1}_3\\\Phi_\R=0\end{array}\right.$};
      \draw (0,-2) node [anchor=north] {$\theta=\frac{3\pi}{2}$, $\left\{\begin{array}{l}\Phi_\L=0\\\Phi_\R=-v\mathds{1}_3\end{array}\right.$};
      \draw [->, thick] (1.8,0) .. controls (1.5,1.2) and (1.2,1.5)  .. (0,1.8);
      \draw (0.9,0.9) node {DW};
    \end{tikzpicture}
    \caption{DW  interpolating the
      $(\Phi_\L,\Phi_\R)=(v\mathds{1}_3,0)$ and the
      $(\Phi_\L,\Phi_\R)=(0,v\mathds{1}_3)$ ground states.
    Increasing the DW profile angle $\theta$ further would interpolate
    to ``parity siblings'' of the first two ground states.
    }
    \label{fig:DW_vacua}
  \end{center}
\end{figure}
Solving the sine-Gordon equation, we have
\beq
\theta = \arctan\exp\left(\pm\sqrt{\frac{\lambda_{34-1-2}}{2}}v(x-x_0)\right)
+ \frac{n\pi}{2},
\label{eq:SG_lambda_sol}
\eeq
which is consistent with the boundary conditions \eqref{eq:BC_theta}
for the upper sign and $n=0$.
There are 3 additional, but equivalent solutions: $n=2$ with the upper
sign, $\varphi_\L\to\varphi_\L+\pi$ and $\varphi_\R\to\varphi_\R+\pi$;
$n=1$ with the lower sign and $\varphi_\L\to\varphi_\L+\pi$ and finally
$n=3$ with the lower sign and $\varphi_\R\to\varphi_\R+\pi$.
We sketch the DW interpolating between the different ground states in
fig.~\ref{fig:DW_vacua}.

Evaluating the total energy of the DW yields
\beq
E = \frac{3\sqrt{\lambda_{34-1-2}}v^3}{\sqrt{2}},
\eeq
where we have set the ground state energy to zero.

We can conclude that in total there are 9 moduli -- 8
orientational parameters of $U(x)$ which is a special unitary matrix as
well as the DW position $x_0$:
\beq
\dim_{\mathbb{R}}\calM^{\gamma=0} = 8+1,
\eeq
where it is understood that this applies to the case without the
$\gamma$ terms turned on.

\subsection{Turning on \texorpdfstring{$\gamma$}{gamma} terms}

We will now turn on the $\gamma_{1,2,3}$ terms of the potential \eqref{eq:V}.
Using the Ansatz \eqref{eq:DWansatz}, the $\gamma$-part of the potential reads
\begin{align}
\calV^{\gamma}&=
\frac{\gamma_1 v^2}{2}\sin(2\theta)\Tr\Big[W^\dag e^{\i(\phi-\psi)}+W e^{-\i(\phi-\psi)}\Big]\non
&\phantom{=\ }
+\frac{\gamma_2 v^4}{4}\sin^2(2\theta)\Tr\left[(W^\dag)^2 e^{\i 2(\phi-\psi)}+W^2 e^{-\i 2(\phi-\psi)}\right]
+\frac{\gamma_3 v^6}{4}\sin^3(2\theta)\cos\left(3(\phi-\psi)\right),
\label{eq:Vgamma}
\end{align}
where we have used $\det W=1$.
First we notice that the $W$, $\phi$ and $\psi$ dependencies do not
drop out of the symmetry-breaking part of the potential, i.e.~of the
$\gamma$-part of the potential.

It is important to notice at this stage, that the vacuum structure has
changed in general.
Although the $\gamma_2$ and $\gamma_3$ terms are minimized in both the
vacua $\theta=0$ and  $\theta=\frac\pi2$, the Josephson ($\gamma_1$)
term is not: Alone it would prefer $\theta=\frac{3\pi}{4}$ or
$\theta=\frac{5\pi}{4}$, but with the $\lambda$ part of the potential
turned on,
its presence simply shifts the vacuum away from the ground states
without the Josephson term.
This has important consequences for the boundary conditions.
The $\gamma_2$ and $\gamma_3$ terms, however, do not shift the vacua
away from $(\Phi_\L,\Phi_\R)=(v\mathds{1}_3,0)$ or
$(\Phi_\L,\Phi_\R)=(0,v\mathds{1}_3)$, nor change the value of
$v$. 
It will thus be convenient to consider the cases with and without the
Josephson term separately.

\subsubsection{Domain walls in Josephson-free chirally broken ground states \texorpdfstring{($\gamma_1=0$, $\gamma_2\leq0$, $\gamma_3\neq0$)}{(gamma1=0, gamma2<=0, gamma3=/=0)}}

Let us start with the simpler case, i.e.~$\gamma_1=0$ and consider the
vacua studied in sec.~\ref{sec:vac_gamma23}.
When the conditions
\eqref{eq:gamma3_vac_lambda12_cond}-\eqref{eq:gamma3_cond} are not
satisfied, the vacuum 
is simply given by eq.~\eqref{eq:vac_gamma3_dw}.

The presence of the $\gamma$ part of the potential \eqref{eq:Vgamma}
with $\gamma_1=0$ changes the equations of motion to 
\begingroup
\allowdisplaybreaks
\begin{align}
  \delta\big(\d W W^\dag\sin^2(2\theta)\big)
  +\gamma_2v^2\sin^2(2\theta)\left(W^2 e^{-\i2(\phi-\psi)} - (W^\dag)^2e^{\i2(\phi-\psi)}\right)
  &=0,\label{eq:eom_gamma_W}\\
  \delta\left(\d\psi\cos^2\theta\right)
  +\frac{\gamma_3 v^4}{8}\sin^3(2\theta)\sin\left(3(\phi-\psi)\right)
  &=0,\label{eq:eom_gamma_psi}\\
  \delta\left(\d\phi\sin^2\theta\right)
  -\frac{\gamma_3 v^4}{8}\sin^3(2\theta)\sin\left(3(\phi-\psi)\right)
  &=0,\label{eq:eom_gamma_phi}\\
  \gamma_2 v^2\sin^2(2\theta)\Tr\left[(W^\dag)^2e^{\i2(\phi-\psi)}-W^2e^{-\i2(\phi-\psi)}\right]&=0,\label{eq:eom_gamma_psiphi}\\
  \delta\d\theta
  +\frac12\left(\d\phi\wedge\star\d\phi-\d\psi\wedge\star\d\psi\right)\sin(2\theta)\non
  \mathop+\frac{v^2}{12}\Tr\left(\frac{\lambda_{34-1-2}}{2}\mathds{1}_3+\frac{\d W^\dag\wedge\star\d W}{v^2}\right)\sin(4\theta)\non
  \mathop+\frac{\gamma_3v^4}{8}\sin(4\theta)\sin(2\theta)\cos\left(3(\phi-\psi)\right)
  &=0,\label{eq:eom_gamma_theta}
\end{align}
\endgroup
where we have taken the trace of eq.~\eqref{eq:eom_gamma_W}, which
implies that eq.~\eqref{eq:eom_gamma_psiphi} holds, which we in
turn have used to simplify the other equations containing this trace.
The equations seem to imply nontrivial functional behavior for $W$,
$\psi$, and $\phi$, but before jumping to conclusions we must
carefully consider the boundary conditions.

Let us now turn to the boundary conditions.
The Ansatz \eqref{eq:DWansatz} is still adequate for the DW with the
$\gamma_{2,3}$ terms turned on, since gauge invariance is not broken by
these terms.
On the other hand, they break chiral symmetry to the diagonal
combination, which has implications for the boundary conditions.
Indeed, since we must now choose $V_\R=V_\L$, the liberty to rotate
the flavors only allows us to diagonalize the boundary condition for
$\Phi_\L$: 
$V_\L V(-\infty)U^{-\infty}V_\L^\dag$ is a diagonal special unitary
matrix.
Since chiral symmetry is broken, we also have $\varphi_\R=\varphi_\L$,
which implies that we cannot independently set the complex phase of
the boundary condition for both $\psi(-\infty)$ and $\phi(\infty)$ to
zero: indeed $\varphi_\L$ is a global transformation.
We thus arrive at the most general boundary conditions for
the case when the $\gamma_{2,3}$ terms are turned on (and $\gamma_1=0$):
\begin{align}
    W(-\infty) &= \diag\left(e^{\i(\alpha+\beta)},e^{\i(-\alpha+\beta)},e^{-\i2\beta}\right), & 
    \phi(\infty) &= 0,\label{eq:BC_gamma_Wphi}\\
\theta(-\infty)&=0,&
\theta(\infty)&=\frac\pi2,\label{eq:BC_gamma_theta}\\
\psi(-\infty)&=\varphi_0,\label{eq:BC_gamma_psi}
\end{align}
which correspond to the fields
\begin{align}
  \Phi_\L(-\infty)&=v\diag\left(e^{\i\alpha_1},e^{\i\alpha_2},e^{\i\alpha_3}\right),&
  \Phi_\L(\infty)&=0,\non
  \Phi_\R(-\infty)&=0,&
  \Phi_\R(\infty)&=v \mathds{1}_3,
\end{align}
where  we have defined
\begin{equation}
\alpha_1:=\varphi_0+\alpha+\beta,\quad
\alpha_2:=\varphi_0-\alpha+\beta,\quad
\alpha_3:=\varphi_0-2\beta,\quad
\varphi_0:=\varphi_U^{-\infty}-\varphi_V^\infty,
\end{equation}
and $\alpha$, $\beta$, $\varphi_0$, $\alpha_{1,2,3}$ are real constants.

Adding together the two equations of motion \eqref{eq:eom_gamma_psi} and
\eqref{eq:eom_gamma_phi}, we get the simple equation 
\beq
\delta(\d\psi\cos^2\theta+\d\phi\sin^2\theta)=0.
\label{eq:eom_reduc_psi_phi_no_gamma1}
\eeq
This equation can be solved by any $\d\psi=\d\phi=c$ with $c$ an
arbitrary real constant (the other possibility of $\d\psi$ being the
inverse of $\cos^2\theta$ would yield a singular solution in the right
vacuum and similarly for $\d\phi$ in the left vacuum).
Although this is a fixed point of the energy, energy minimization
selects $c=0$ rendering $\psi$ and $\phi$ constant.
Using now the boundary conditions \eqref{eq:BC_gamma_psi} and
\eqref{eq:BC_gamma_Wphi}, we can conclude that $\psi=\varphi_0$ and
$\phi=0$ for all $x$.
This is a great simplification of the equations of motion.
Using now that $\d\psi=\d\phi=0$, eqs.~\eqref{eq:eom_gamma_psi} and
\eqref{eq:eom_gamma_phi} tell us that an arbitrary boundary condition
$\varphi_0$ does not solve the equations of motion (with a smooth
solution for the phase functions $\psi(x)$ and $\phi(x)$) and we must
select the boundary conditions as
\beq
\varphi_0=\frac{n\pi}{3}, \qquad n\in\mathbb{Z}.
\eeq
Since $\psi$ only possesses a boundary condition in the left-vacuum
($\psi(-\infty)$), there is no physical distinction between the
different choices of $n$ in the above equation; we may as well select
$n=0$ for simplicity.
We can now turn to the algebraic equation \eqref{eq:eom_gamma_psiphi},
where $\sin(2\theta)$ cannot vanish in the DW solution and we assume
$\gamma_2\neq 0$; this leads to
\beq
\Tr\big[(W^\dag)^2-W^2\big] = 0.
\label{eq:W_algebraic}
\eeq
We do not have a rigorous proof that $W$ must be a constant matrix 
everywhere, although that is both consistent with the boundary
conditions \eqref{eq:BC_Wphi} and minimizes its kinetic term.
Indeed from eq.~\eqref{eq:eom_gamma_theta}, we can see that a
nonconstant $W$ can only increase the DW tension and hence a constant
$W$ minimizes the DW tension.
A constant solution for $W$ is easy to find, as it is simply given by
its boundary condition \eqref{eq:BC_gamma_Wphi} and the condition
\eqref{eq:W_algebraic} yields only two solutions: $\beta=n\pi/2$ or
$\alpha=\pm\beta+\pi m$, ($n,m\in\mathbb{Z}$),
which have the form
\begin{align}
W &= \diag\left(\pm e^{\i\alpha},e^{-\i\alpha},\pm1\right), \qquad
\textrm{or}\non
W &= \diag\left(\pm e^{\i2\beta},\pm 1,e^{-\i2\beta}\right), \qquad
\textrm{or}\non
W &= \diag\left(\pm 1,\pm e^{\i2\beta},e^{-\i2\beta}\right),
\label{eq:Wsols}
\end{align}
which are in total 6 solutions that take into account all
possibilities of shifts with $n$ and $m$. 

Having solved the equations of motion for $W$, $\psi$ and $\phi$,
let us reduce the $\gamma$-part of the potential\footnote{One may
worry whether the change of the boundary condition will affect the
vacua. This is not the case for the DW vacua \eqref{eq:vac_gamma3_dw},
but the conditions for the vacua's existence
\eqref{eq:gamma3_vac_lambda12_cond}-\eqref{eq:gamma3_cond} 
should be evaluated with $3\gamma_2\to(1+2\cos(2\alpha))\gamma_2$.
For perturbatively small $\gamma_2$, this does not affect the vacuum
properties.
}:
\beq
\calV^{\gamma}=
\frac{\gamma_2 v^4}{2}\big(1+2\cos(2\alpha)\big)\sin^2(2\theta)
+\frac{\gamma_3 v^6}{4}\sin^3(2\theta),
\eeq
where for the boundary condition with $\beta$ in eq.~\eqref{eq:Wsols},
we have replaced $\beta\to\frac\alpha2$.
Interestingly enough, for certain boundary conditions the
Josephson-squared ($\gamma_2$) term vanishes:
\beq
\alpha=\frac{\pi}{3}, \qquad \textrm{or}\qquad
\alpha=\frac{2\pi}{3}, \qquad \textrm{or}\qquad
\alpha=\frac{4\pi}{3}, \qquad \textrm{or}\qquad
\alpha=\frac{5\pi}{3},
\eeq
where we assumed $\alpha$ to be in the range from 0 to $2\pi$.
The $\gamma$-part of the potential thus depends strongly on the choice
of boundary conditions for $W$.
Notice though that the $\gamma_3$ term is insensitive to the choice of
boundary conditions. 

Let us turn to the equation for the DW profile, namely that for
$\theta$:
\beq
\delta\d\theta
  +\gamma_2'\sin(4\theta)
  +\frac32\gamma_3'\sin(4\theta)\sin(2\theta)
  &=0,\label{eq:eom_gamma_theta_simpl}
\eeq
where we have defined
\begin{align}
  \gamma_2'&=\frac{v^2}{2}\left(\frac{\lambda_{34-1-2}}{4} + \frac{\gamma_2}{3}\big(1+2\cos(2\alpha)\big)\right),\\
  \gamma_3'&=\frac{\gamma_3v^4}{12}.
\end{align}
This equation is not integrable to the best of our knowledge (with
both $\gamma_2'\neq0$ and $\gamma_3'\neq0$).
We will assume that the chiral symmetry breaking parameters are small
compared to the other parameters in the GL EFT potential \eqref{eq:V}
and hence that $\lambda_{34-1-2}>4\gamma_2$.
The vacuum condition necessary for DWs tells us that
$\lambda_{34-1-2}>0$ and with the latter assumption, we have that
$\gamma_2'>0$ as well.
Indeed this guarantees a positive mass term for the sine-Gordon
equation and hence a DW, at least in the absence of the $\gamma_3'$
term.
Turning on $\gamma_3'$ modifies the DW from the standard sine-Gordon
solution.
Since $\gamma_2\leq0$, a nonvanishing negative value of the $\gamma_2$ will increase the DW tension for $\alpha\in\big[\tfrac{\pi}{3},\tfrac{2\pi}{3}\big]$ or
$\alpha\in\big[\tfrac{4\pi}{3},\tfrac{5\pi}{3}\big]$ and decreases it otherwise.

\subsubsection{Josephson chirally broken ground states
  \texorpdfstring{($\gamma_1\neq0$, $\gamma_2\in\mathbb{R}$, $\gamma_3=0$)}{(gamma1=/=0, gamma2 in R, gamma3=0)}: Non\-existence of domain walls}

Let us turn to the case of a nonvanishing Josephson term
($\gamma_1\neq0$); for simplicity, we here allow $\gamma_2\neq 0$ but
keep $\gamma_3=0$ turned off.
This case is complicated by the fact that the vacuum properties are
changed, although for small $\gamma_1$, only perturbatively, but
mathematically $\Phi_\R=0$ or $\Phi_\R\neq0$ in the
$\Phi_\L\propto\mathds{1}_3$ vacuum makes a big difference when using
global transformations to reduce the complexity of the problem.

The vacua with $\gamma_1\neq0$, $\gamma_3=0$ and $\gamma_2$ can be
turned on or left switched off, is given in
eq.~\eqref{eq:vac_gamma1_dw}.
Crucially none of the fields vanish for $\gamma_1\neq0$.
We now need to modify the Ansatz for the domain wall.
Imposing still the nonlinear sigma-model limit
\eqref{eq:sigma_model_limit}, we can thus write
\begin{align}
\Phi_\L&=\big(w_+\cos\theta(x)-w_-\sin\theta(x)\big)U(x)e^{\i\varphi_U(x)},\non
\Phi_\R&=\big(w_+\sin\theta(x)+w_-\cos\theta(x)\big)V(x)e^{\i\varphi_V(x)}.
\end{align}
Using the boundary conditions
\begin{align}
U(-\infty)&=U^{-\infty},& 
U(\infty)&=V^\infty,\\
V(-\infty)&=U^{-\infty},&
V(\infty)&=V^\infty,\\
\varphi_U(-\infty)&=\varphi_U^{-\infty},&
\varphi_U(\infty)&=\varphi_V^{\infty}\\
\varphi_V(-\infty)&=\varphi_U^{-\infty}+\frac{\pi}{2}(1+\sign(\gamma_1)),&
\varphi_V(\infty)&=\varphi_V^{\infty}+\frac{\pi}{2}(1-\sign(\gamma_1))\\
\theta(-\infty)&=0,&
\theta(\infty)&=\frac\pi2,
\end{align}
the fields in the two vacua read
\begin{align}
  \Phi_\L(-\infty)&=w_+ U^{-\infty} e^{\i\varphi_U^{-\infty}},&
  \Phi_\L(\infty)&=-w_- V^\infty e^{\i\varphi_V^\infty},\non
  \Phi_\R(-\infty)&=-\sign(\gamma_1) w_- U^{-\infty} e^{\i\varphi_U^{-\infty}},&
  \Phi_\R(\infty)&=\sign(\gamma_1) w_+ V^\infty e^{\i\varphi_V^\infty}.
\end{align}
We notice a doubling of nonvanishing boundary conditions with respect to
eqs.~\eqref{eq:BC_pre1}-\eqref{eq:BC_pre3} (except for $\theta$ that
retains its boundary conditions).

We should comment on the different choice of boundary condition here,
as compared with the previous cases.
Indeed, we can no longer parametrize the vacuum with two independent
unitary matrices, since the vacuum equation has a nonvanishing
contribution from the Josephson term $\Tr[\Phi_\L^\dag\Phi_\R]+{\rm c.c.}$
This is the reason for the matched choice of the boundary conditions
$U(-\infty)=V(-\infty)=U^{-\infty}$ as well as
$U(\infty)=V(\infty)=V^\infty$.
The same token goes for the complex phases, except for taking into
account the appropriate relative signs of $\Phi_\L$ and $\Phi_\R$ that
depends on the sign of $\gamma_1$, see eq.~\eqref{eq:vac_gamma1_dw}.

Using all transformations at once in this case, now yields
\begin{align}
\Phi_\L&=\big(w_+\cos\theta(x)-w_-\sin\theta(x)\big)g_C(x)U(x)V_\L^\dag e^{\i(\varphi_U(x)+\varphi_\L)},\non
\Phi_\R&=\big(w_+\sin\theta(x)+w_-\cos\theta(x)\big)g_C(x)V(x)V_\L^\dag e^{\i(\varphi_V(x)+\varphi_\L)},
\end{align}
where we have used the fact that in the chirally broken phase $V_\R=V_\L$ and $\varphi_\R=\varphi_\L$. 
Choosing to simplify the right field, we perform the gauge transformation
\beq
g_C(x) = V_\L V(x)^\dag,
\eeq
which gives
\begin{align}
\Phi_\L&=\big(w_+\cos\theta(x)-w_-\sin\theta(x)\big)V_\L V(x)^\dag U(x)V_\L^\dag e^{\i(\varphi_U(x)+\varphi_\L)},\non
\Phi_\R&=\big(w_+\sin\theta(x)+w_-\cos\theta(x)\big)\mathds{1}_3 e^{\i(\varphi_V(x)+\varphi_\L)}.
\end{align}
Looking now at the vacua, we have
\begin{align}
  \Phi_\L(-\infty)&=w_+ \mathds{1}_3 e^{\i\varphi_0},&
  \Phi_\L(\infty)&=-w_- \mathds{1}_3,\non
  \Phi_\R(-\infty)&=-\sign(\gamma_1) w_- \mathds{1}_3 e^{\i\varphi_0},&
  \Phi_\R(\infty)&=\sign(\gamma_1) w_+ \mathds{1}_3,
\end{align}
where we have chosen $\varphi_\L=-\varphi_V^\infty$ and defined $\varphi_0:=\varphi_U^{-\infty}-\varphi_V^\infty$.

We can now define the field $W(x):=V_\L V(x)^\dag U(x)V_\L^\dag$, as
before, and write the DW Ansatz in the final form:
\begin{align}
\Phi_\L&=\big(w_+\cos\theta(x)-w_-\sin\theta(x)\big)e^{\i\psi(x)} W(x),\non
\Phi_\R&=\big(w_+\sin\theta(x)+w_-\cos\theta(x)\big)e^{\i\phi(x)}\mathds{1}_3,
\label{eq:DWansatz_gamma1}
\end{align}
with the boundary conditions
\begin{align}
  W(-\infty)&=\mathds{1}_3,&
  W(\infty)&=\mathds{1}_3,\label{eq:BC_gamma1_W}\\
  \psi(-\infty)&=\varphi_0,&
  \psi(\infty)&=0,\label{eq:BC_gamma1_psi}\\
  \phi(-\infty)&=\varphi_0+\frac{\pi}{2}(1+\sign(\gamma_1)),&
  \phi(\infty)&=\frac{\pi}{2}(1-\sign(\gamma_1)),\label{eq:BC_gamma1_phi}\\
  \theta(-\infty)&=0,&\theta(\infty)&=\frac\pi2.\label{eq:BC_gamma1_theta}
\end{align}
We cannot use symmetry considerations to set the complex phase
$\varphi_0$ to zero, which in general turns on nontrivial behavior
for both phase functions $\psi$ and $\phi$.
Notice that no $\varphi_0$ exists that will allow for both a constant
solution for $\psi$ and $\phi$.
This implies that one of the functions, $\psi$ or $\phi$ (or both),
must be nontrivial in order to obey the above boundary conditions.

Recalculating the $\SU(3)$ gauge field in the sigma-model limit, we
obtain
\beq
A = \frac{(w_+\cos\theta-w_-\sin\theta)^2}{w_+^2+w_-^2}\d W W^{-1},
\eeq
and we have 
\begin{align}
  w_+^2+w_-^2&=\frac{m^2}{\lambda_{12}}=v^2,\\
  w_+^2-w_-^2&=\frac{m^2}{\lambda_{12}}\sqrt{1-\frac{16\gamma_1^2\lambda_{12}^2}{m^4(\lambda_{34-1-2}+4\gamma_2)^2}},\\
  w_+w_-&=\frac{2|\gamma_1|}{|\lambda_{34-1-2}+4\gamma_2|},
\end{align}
for reference.
The energy functional in the sigma-model limit with the Ansatz \eqref{eq:DWansatz_gamma1} thus reads
\begin{align}
  E&= \frac{1}{4(w_+^2+w_-^2)}\left\|\big((w_+^2-w_-^2)\sin(2\theta)+2w_+w_-\cos(2\theta)\big)\d W\right\|^2\\
  &\phantom{=\ }
  +\|(w_+\cos\theta-w_-\sin\theta)\d\psi\mathds{1}_3\|^2
  +\|(w_+\sin\theta+w_-\cos\theta)\d\phi\mathds{1}_3\|^2
  +(w_+^2+w_-^2)\|\d\theta\mathds{1}_3\|^2\non
  &\phantom{=\ }
  +\frac{\lambda_{34-1-2}}{8}\left\|\big((w_+^2-w_-^2)\sin(2\theta)+2w_+w_-\cos(2\theta)\big)\mathds{1}_3\right\|^2\non
  &\phantom{=\ }
  +\frac{\gamma_1}{2}\int_M\star\big((w_+^2-w_-^2)\sin(2\theta)+2w_+w_-\cos(2\theta)\big)\Tr\left[W^\dag e^{\i(\phi-\psi)} + W e^{-\i(\phi-\psi)}\right]\non
  &\phantom{=\ }
  +\frac{\gamma_2}{4}\int_M\star\big((w_+^2-w_-^2)\sin(2\theta)+2w_+w_-\cos(2\theta)\big)^2\Tr\left[(W^\dag)^2 e^{\i2(\phi-\psi)} + W^2 e^{-\i2(\phi-\psi)}\right].\nonumber
\end{align}
In the limit $\gamma_1\to0$, the VEVs simplify as $w_+\to v$ and
$w_-\to0$, which in turn simplify the above energy functional to that
of eq.~\eqref{eq:Esigmamodel_lambda} for the non-$\gamma$ part and the
Josephson-squared term reduces to that in eq.~\eqref{eq:Vgamma}. 
The equations of motion corresponding to the above energy functional
are given by
\begingroup
\allowdisplaybreaks
\begin{align}
  \frac{1}{w_+^2+w_-^2}\delta\left(\big((w_+^2-w_-^2)\sin(2\theta)+2w_+w_-\cos(2\theta)\big)^2\d W W^\dag\right)\non
  \mathop+\gamma_1\big((w_+^2-w_-^2)\sin(2\theta)+2w_+w_-\cos(2\theta)\big)\left(W e^{-\i(\phi-\psi)} - W^\dag e^{\i(\phi-\psi)}\right)\non
  \mathop+\gamma_2\big((w_+^2-w_-^2)\sin(2\theta)+2w_+w_-\cos(2\theta)\big)^2\left(W^2 e^{-\i2(\phi-\psi)} - (W^\dag)^2 e^{\i2(\phi-\psi)}\right)
  &=0,\label{eq:eom_gamma1_W}\\
  \delta\left((w_+\cos\theta-w_-\sin\theta)^2\d\psi\right)&=0,\label{eq:eom_gamma1_psi}\\
  \delta\left((w_+\sin\theta+w_-\cos\theta)^2\d\phi\right)&=0,\label{eq:eom_gamma1_phi}\\
  \gamma_1\big((w_+^2-w_-^2)\sin(2\theta)+2w_+w_-\cos(2\theta)\big)\Tr\left[W e^{-\i(\phi-\psi)} - W^\dag e^{\i(\phi-\psi)}\right]\non
  \mathop+\gamma_2\big((w_+^2-w_-^2)\sin(2\theta)+2w_+w_-\cos(2\theta)\big)^2\Tr\left[W^2 e^{-\i2(\phi-\psi)} - (W^\dag)^2 e^{\i2(\phi-\psi)}\right]&=0,\label{eq:eom_gamma1_psiphi}\\
  \delta(\d\theta)
  +\frac{(w_+^4+w_-^4-6w_+^2w_-^2)\sin(4\theta) + 4(w_+^2-w_-^2)w_+w_-\cos(4\theta)}{12(w_+^2+w_-^2)^2}\Tr\big[\d W^\dag\d W\big]\non
  \mathop+\frac{(w_+^2-w_-^2)\sin(2\theta)+2w_+w_-\cos(2\theta)}{2(w_+^2+w_-^2)}\left(\d\phi\wedge\star\d\phi-\d\psi\wedge\star\d\psi\right)\non
  \mathop+\lambda_{34-1-2}\frac{(w_+^4+w_-^4-6w_+^2w_-^2)\sin(4\theta) + 4(w_+^2-w_-^2)w_+w_-\cos(4\theta)}{8(w_+^2+w_-^2)}\non
  \mathop+\gamma_1\frac{(w_+^2-w_-^2)\cos(2\theta)-2w_+w_-\sin(2\theta)}{6(w_+^2+w_-^2)}\Tr\left[W^\dag e^{\i(\phi-\psi)} + W e^{-\i(\phi-\psi)}\right]\non
  \mathop+\gamma_2\frac{(w_+^4+w_-^4-6w_+^2w_-^2)\sin(4\theta) + 4(w_+^2-w_-^2)w_+w_-\cos(4\theta)}{12(w_+^2+w_-^2)}\non\times\Tr\left[(W^\dag)^2 e^{\i2(\phi-\psi)} + W^2 e^{-\i2(\phi-\psi)}\right]&=0,\label{eq:eom_gamma1_theta}
\end{align}
\endgroup
where we have used that the trace-part of eq.~\eqref{eq:eom_gamma1_W}
implies eq.~\eqref{eq:eom_gamma1_psiphi} and in turn reduces
eqs.~\eqref{eq:eom_gamma1_psi} and \eqref{eq:eom_gamma1_phi} to
first-order differential equations.
Since we have defined $w_\pm>0$, i.e.~to be positive, we can conclude
from eq.~\eqref{eq:eom_gamma1_psi} that $\d\psi=0$ when
$\theta\in\big[0,\tfrac\pi2\big]$ and hence that
$\psi$ is a constant for all $x$.

We can already now see the boundary conditions are spelling trouble
for the existence of the solution to the DW.
Indeed, if we do not choose $\varphi_0=0$, we cannot have a solution
with $\psi$ possessing an everywhere smooth derivative.
So let us for now choose $\varphi_0:=0$.

The next issue on the horizon is the necessity of the nontrivial
$\phi$ behavior.
Indeed, the boundary conditions \eqref{eq:BC_gamma1_phi} dictate that
$\phi(-\infty)=\pi$ ($\phi(-\infty)=0$) and $\phi(\infty)=0$
($\phi(\infty)=\pi$) for $\gamma_1>0$ ($\gamma_1<0$).
A constant solution is thus not an option in either case.

We will now establish the following theorem of nonexistence.
\begin{theorem}
  There exist no smooth regular solutions for $\phi(x)$, $\psi(x)$ and $W(x)$ for
  $\gamma_1\neq0$, $\gamma_2=0$ and $w_+>w_-$ that obey the boundary conditions
  \eqref{eq:BC_gamma1_W}-\eqref{eq:BC_gamma1_phi}.
  \label{thm:1}
\end{theorem}
\emph{Proof}:
We recall that $\varphi_0=0$ is necessary for the equation
\eqref{eq:eom_gamma1_psi} to be solved by a constant solution,
i.e.~$\psi(x)=0$.
Indeed, if we look for a nontrivial solution for $\psi(x)$, the first integral
of its equation of motion yields a pole in the derivative:
\beq
\frac{\d\psi}{\d x}=\frac{A_\psi}{(w_+\cos\theta(x)-w_-\sin\theta(x))^2},
\eeq
where we recall that $w_\pm>0$ by definition of the vacuum equations.
We thus choose $\varphi_0:=0$ and proceed.
Integrating similarly the equation \eqref{eq:eom_gamma1_phi}, we have
\beq
\frac{\d\phi}{\d x}=\frac{A_\phi}{(w_+\sin\theta(x)+w_-\cos\theta(x))^2},
\label{eq:phi_derivative}
\eeq
which does not contain a pole.
We will now assume that $\theta(x)$ is a DW-type solution that
approaches its boundary condition \eqref{eq:BC_gamma1_theta}
exponentially
\beq
\lim_{x\to-\infty}\theta(x)=0+\mathcal{O}(e^{-C x}),
\eeq
for $C>0$ a positive constant.
We can therefore infer that the derivative,
\beq
\frac{\d\phi}{\d x}\approx\frac{A_\phi}{w_-^2},
\eeq
is nonvanishing all the way
to $x\to-\infty$, unless we set $A_\phi:=0$.
Setting $A_\phi:=0$, however, does not yield a nontrivial solution
that can interpolate $\phi(x)$ between $\pi$ and $0$ (or $0$ and
$\pi$, depending on the sign of $\gamma_1$).
Let us entertain the possibility of an unorthodox solution, where $\phi$ is winding
continuously from $x\to-\infty$ to the DW.
We then inspect eq.~\eqref{eq:eom_gamma1_psiphi}, which for
$\gamma_2:=0$ depends only on the trace of $W$ and its Hermitian
conjugate.
We can thus assume the Ansatz for $W$ (since the equation in question
does not depend on the non-trace part of $W$):
\beq
W=\diag\left(e^{\i(\alpha+\beta)},e^{\i(-\alpha+\beta)},e^{-\i2\beta}\right),
\eeq
for which eq.~\eqref{eq:eom_gamma1_psiphi} reads
\begin{align}
\sin\phi\left(\cos(\alpha+\beta)+\cos(\alpha-\beta)+\cos(2\beta)\right)\non
+\cos\phi\left(-\sin(\alpha+\beta)+\sin(\alpha-\beta)+\sin(2\beta)\right)&=0,
\end{align}
where we have used that neither $\gamma_1$ nor the function 
$\big((w_+^2-w_-^2)\sin(2\theta)+2w_+w_-\cos(2\theta)\big)$ vanishes.
In particular, the latter function is nonzero in the vacuum limit
$\theta\to0$ or $\theta\to\pi$ (we recall that $\gamma_1\neq 0$
implies that $w_-\neq0$, see eq.~\eqref{eq:vac_gamma1_dw}).
In the left vacuum ($x\to-\infty$) we have
$\phi=\frac{\pi}{2}(1+\sign(\gamma_1))$, for which $\sin\phi=0$, so to
satisfy the equation, we set $\beta:=0$.
For $\phi$ to reach the other boundary condition at $x\to\infty$,
i.e.~$\phi=\frac{\pi}{2}(1-\sign(\gamma_1))$ it must deviate from its
vacuum at $x\to-\infty$ for which $\sin\phi$ no longer vanishes.
This requires us to make $2\cos\alpha+1$ vanish:
\beq
\alpha=\frac{2\pi}{3},\qquad \textrm{or}\qquad
\alpha=\frac{4\pi}{3}.
\label{eq:Wsol_gamma1_alpha}
\eeq
Comparing with eq.~\eqref{eq:phi_derivative}, we observe that $\phi$'s
derivative is everywhere nonvanishing, so $\sin\phi$ cannot vanish,
except for a Lebesgue measure zero subset of the real line.
Inserting now the constant solution for $W$ into its equation of
motion \eqref{eq:eom_gamma1_W}, it is clear that although the trace of
$W e^{-\i(\phi-\psi)} - W^\dag e^{\i(\phi-\psi)}$ vanishes, the matrix
clearly does not.
This means that the derivative of $W$ cannot vanish, as we have
already established that
$\big((w_+^2-w_-^2)\sin(2\theta)+2w_+w_-\cos(2\theta)\big)$ does not
vanish, even in the vacuum.
Finally, the constant solution with $\alpha$ given in
eq.~\eqref{eq:Wsol_gamma1_alpha} is not compatible with the boundary
condition \eqref{eq:BC_gamma1_W}.
\hfill$\square$

\subsubsection{Josephson chirally broken ground states
  \texorpdfstring{($\gamma_1\neq0$, $\gamma_2\in\mathbb{R}$, $\gamma_3=0$)}{(gamma1=/=0, gamma2 in R, gamma3=0)}: Kink}

\begin{figure}[!ht]
  \begin{center}
    \begin{tikzpicture}[scale=1.0]
      \draw (0,0) circle (2);
      \filldraw [gray] (2,0) circle (0.1);
      \draw [thick] (-0.1,1.9) -- (0.1,2.1);
      \draw [thick] (-0.1,2.1) -- (0.1,1.9);
      \filldraw [gray] (-2,0) circle (0.1);
      \draw [thick] (-0.1,-1.9) -- (0.1,-2.1);
      \draw [thick] (-0.1,-2.1) -- (0.1,-1.9);
      \draw (2,0) node [right] {$\ \theta=0$, $\left\{\begin{array}{l}\Phi_\L=w_+\mathds{1}_3\\\Phi_\R=-\sign(\gamma_1)w_-\mathds{1}_3\end{array}\right.$};
      \draw (0,2) node [anchor=south] {$\theta=\frac{\pi}{2}$, $\left\{\begin{array}{l}\Phi_\L=-w_-\mathds{1}_3\\\Phi_\R=\sign(\gamma_1)w_+\mathds{1}_3\end{array}\right.$};
      \draw (-2,0) node [left] {$\theta=\pi$, $\left\{\begin{array}{l}\Phi_\L=-w_+\mathds{1}_3\\\Phi_\R=\sign(\gamma_1)w_-\mathds{1}_3\end{array}\right.$};
      \draw (0,-2) node [anchor=north] {$\theta=\frac{3\pi}{2}$, $\left\{\begin{array}{l}\Phi_\L=w_-\mathds{1}_3\\\Phi_\R=-\sign(\gamma_1)w_+\mathds{1}_3\end{array}\right.$};
      \draw [->, thick] (1.8,0) .. controls (1.5,2.3) and (-1.5,2.3)  .. (-1.8,0);
      \draw (0.9,0.9) node {kink};
    \end{tikzpicture}
    \caption{The kink soliton interpolates between the vacuum at
      $\theta=0$ and the sign-flipped vacuum at $\theta=\pi$.
    It is fundamentally different from the DW solution that swaps the
    two vacua (up to an overall sign for negative $\gamma_1$), which
    would correspond to a solution that interpolates between the
    ground state at $\theta=0$ and the would-be ground state at
    $\theta=\frac\pi2$.
    Instead, the DW solution without the correct signs at $\frac\pi2$,
    i.e.~a solution that goes to $\Phi_\L=-w_-\mathds{1}_3$,
    $\Phi_\R=-\sign(\gamma_1)w_+\mathds{1}_3$ exists, but that is not
    a ground state.
    The ground states in the figure at $\theta=\frac\pi2$ and
    $\frac{3\pi}{2}$ are thus marked with $\times$, whereas the true
    ground states that are reachable by the kink are marked with
    gray-filled circles.
    }
    \label{fig:Josephson_vacua}
  \end{center}
\end{figure}
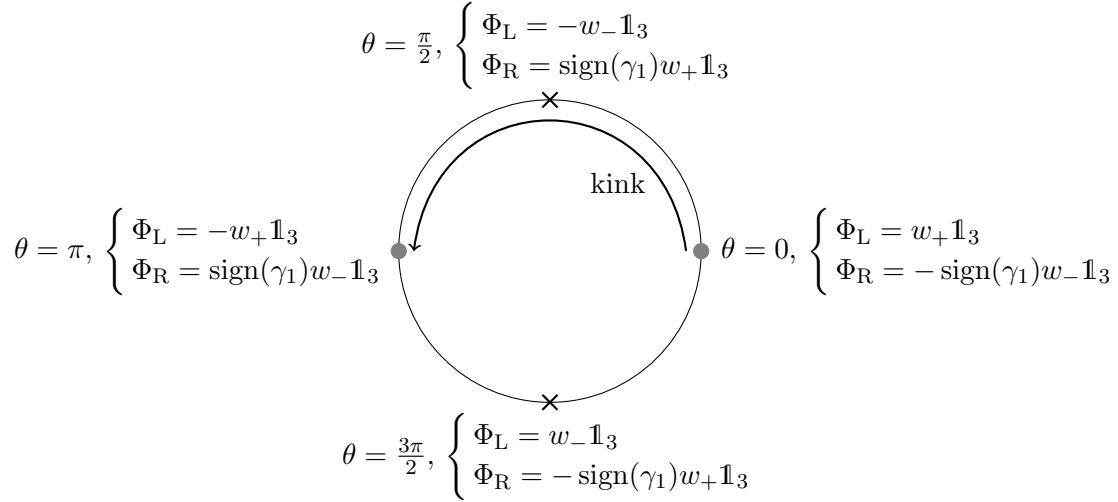

Although the minimal DW interpolating between the two different vacua
does not exist, see fig.~\ref{fig:Josephson_vacua}, there is still the
possibility of the soliton going from the Josephson vacuum to its
parity sibling (i.e.~the same ground state, but with the signs of both
$\Phi_\L$ and $\Phi_\R$ flipped).
We denote this soliton a kink due to its resemblance with the
sine-Gordon kink.

We will now change the boundary conditions for $\phi$ and $\theta$ to:
\begin{align}
  W(-\infty)&=\mathds{1}_3,&
  W(\infty)&=\mathds{1}_3,\label{eq:BC_gamma1_W_kink}\\
  \psi(-\infty)&=\varphi_0,&
  \psi(\infty)&=0,\label{eq:BC_gamma1_psi_kink}\\
  \phi(-\infty)&=\varphi_0+\frac{\pi}{2}(1+\sign(\gamma_1)),&
  \phi(\infty)&=\frac{\pi}{2}(1+\sign(\gamma_1)),\label{eq:BC_gamma1_phi_kink}\\
  \theta(-\infty)&=0,&\theta(\infty)&=\pi,\label{eq:BC_gamma1_theta_kink}
\end{align}
whereas the equations of motion remain those of
eqs.~\eqref{eq:eom_gamma1_W}-\eqref{eq:eom_gamma1_theta} and the
boundary conditions for $W$ and $\psi$ are unchanged with respect to
eqs.~\eqref{eq:BC_gamma1_W} and \eqref{eq:BC_gamma1_psi}.
These boundary conditions correspond to the vacua
\begin{align}
  \Phi_\L(-\infty)&=w_+ \mathds{1}_3 e^{\i\varphi_0},&
  \Phi_\L(\infty)&=-w_+ \mathds{1}_3,\non
  \Phi_\R(-\infty)&=-\sign(\gamma_1) w_- \mathds{1}_3 e^{\i\varphi_0},&
  \Phi_\R(\infty)&=\sign(\gamma_1) w_- \mathds{1}_3,
\end{align}
which are indeed the same ground states on the left and on the right
of the kink, up to the overall sign flip.

We do not have a rigorous proof that $W$ should be a constant for all
$x$, but energy minimization suggests that this is the solution.
The boundary conditions \eqref{eq:BC_gamma1_W_kink} then determines
$W=\mathds{1}_3$ everywhere.
If we assume this to be the case, then
eq.~\eqref{eq:eom_gamma1_psiphi} reduces to
\beq
\sin(\phi-\psi)\left[
  \gamma_1
  +4\gamma_2(w_+\cos\theta-w_-\sin\theta)(w_+\sin\theta+w_-\cos\theta)\cos(\phi-\psi)\right]=0.
\label{eq:eom_gamma1_psiphi_reduc}
\eeq
Now since $(w_+\cos\theta-w_-\sin\theta)$ vanishes\footnote{If we flip
the sign of $w_-$, the other parenthesis vanishes. } no solution of
$\phi-\psi$ can make the cosine term cancel $\gamma_1$.
This means that the only solution to this equation is
$\phi-\psi=n\pi$, $n\in\mathbb{Z}$.
Since $\psi$ must be a constant, then so must $\phi$.
We observe that the parameter, $\varphi_0$, of the boundary conditions
\eqref{eq:BC_gamma1_psi} and \eqref{eq:BC_gamma1_phi} drops out of the
combination $\phi-\psi$ and that they determine
\beq
\phi-\psi=\frac{\pi}{2}(1+\sign(\gamma_1)),
\eeq
which is $\pi$ ($0$) for positive (negative) Josephson coupling,
$\gamma_1$.
Luckily, both are solutions to
eq.~\eqref{eq:eom_gamma1_psiphi_reduc}.
This solution for $\phi-\psi$ solves the equation of motion for
$W$, provided $W=\mathds{1}_3$.

Using these solutions, the equations of motion
\eqref{eq:eom_gamma_W}-\eqref{eq:eom_gamma_theta} reduce to just:
\beq
\delta\d\theta
+\gamma_1'\sin(2\theta)
+\gamma_1''\cos(2\theta)
+\gamma_2'\sin(4\theta)
+\gamma_2''\cos(4\theta)
  &=0,\label{eq:eom_gamma1_theta_simpl}
\eeq
where we have defined
\begin{align}
  \gamma_1'&=\frac{2|\gamma_1|w_+w_-}{w_+^2+w_-^2}
  =\frac{4\gamma_1^2|\lambda_{12}|}{m^2|\lambda_{34-1-2}+4\gamma_2|},\\
  \gamma_1''&=-\frac{|\gamma_1|(w_+^2-w_-^2)}{w_+^2+w_-^2}
  =-|\gamma_1|\sqrt{1-\frac{16\gamma_1^2\lambda_{12}^2}{m^4(\lambda_{34-1-2}+4\gamma_2)^2}},\\
  \gamma_2'&=\frac{(\lambda_{34-1-2}+4\gamma_2)(w_+^4+w_-^4-6w_+^2w_-^2)}{8(w_+^2+w_-^2)}
  =\frac{m^2(\lambda_{34-1-2}+4\gamma_2)}{8\lambda_{12}}-\frac{4\gamma_1^2\lambda_{12}}{m^2(\lambda_{34-1-2}+4\gamma_2)},\\
  \gamma_2''&=\frac{(\lambda_{34-1-2}+4\gamma_2)(w_+^2-w_-^2)w_+w_-}{2(w_+^2+w_-^2)}
  =|\gamma_1|\sqrt{1-\frac{16\gamma_1^2\lambda_{12}^2}{m^4(\lambda_{34-1-2}+4\gamma_2)^2}}.
\end{align}
Equation \eqref{eq:eom_gamma1_theta_simpl} is expected on general
grounds to have analytic solutions of the form of an elliptic function
of arcsin of a root to a polynomial equation, which in practice can be
quite cumbersome to work out explicitly.

As a check, we can take the two limits, $\theta\to0$ and
$\theta\to\pi$, of eq.~\eqref{eq:eom_gamma1_theta_simpl}; in both
cases, the non-derivative terms are given by:
\beq
\gamma_1''+\gamma_2'' = 0,
\eeq
which vanishes due to the ground state equations.
It is also straightforward to check that $\theta\to\tfrac\pi2$ does
not vanish; the non-derivative terms become:
\beq
-\gamma_1''+\gamma_2''
= 2|\gamma_1|\sqrt{1-\frac{16\gamma_1^2\lambda_{12}^2}{m^4(\lambda_{34-1-2}+4\gamma_2)^2}}
\neq 0,
\eeq
except for the fine-tuned point in the theory where $w_+=w_-$, see
the next section.

\subsubsection{Josephson chirally broken ground states
  \texorpdfstring{($\gamma_1\neq0$, $\gamma_2\in\mathbb{R}$, $\gamma_3=0$)}{(gamma1=/=0, gamma2 in R, gamma3=0)}: Kink at fine-tuned point}\label{sec:DW_fine_tuned}

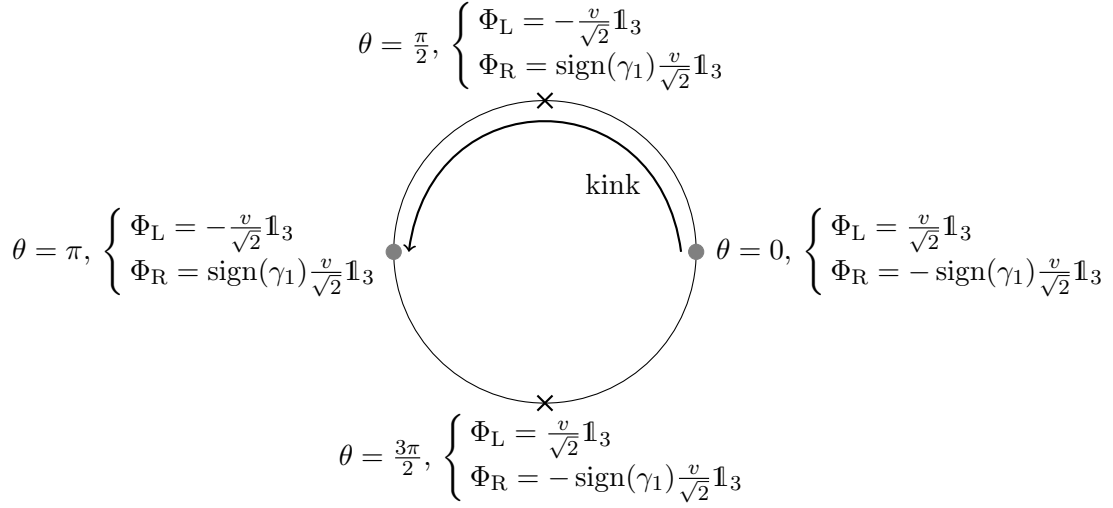
\begin{figure}[!ht]
  \begin{center}
    \begin{tikzpicture}[scale=1.0]
      \draw (0,0) circle (2);
      \filldraw [gray] (2,0) circle (0.1);
      \draw [thick] (-0.1,1.9) -- (0.1,2.1);
      \draw [thick] (-0.1,2.1) -- (0.1,1.9);
      \filldraw [gray] (-2,0) circle (0.1);
      \draw [thick] (-0.1,-1.9) -- (0.1,-2.1);
      \draw [thick] (-0.1,-2.1) -- (0.1,-1.9);
      \draw (2,0) node [right] {$\ \theta=0$, $\left\{\begin{array}{l}\Phi_\L=\tfrac{v}{\sqrt{2}}\mathds{1}_3\\\Phi_\R=-\sign(\gamma_1)\tfrac{v}{\sqrt{2}}\mathds{1}_3\end{array}\right.$};
      \draw (0,2) node [anchor=south] {$\theta=\frac{\pi}{2}$, $\left\{\begin{array}{l}\Phi_\L=-\tfrac{v}{\sqrt{2}}\mathds{1}_3\\\Phi_\R=\sign(\gamma_1)\tfrac{v}{\sqrt{2}}\mathds{1}_3\end{array}\right.$};
      \draw (-2,0) node [left] {$\theta=\pi$, $\left\{\begin{array}{l}\Phi_\L=-\tfrac{v}{\sqrt{2}}\mathds{1}_3\\\Phi_\R=\sign(\gamma_1)\tfrac{v}{\sqrt{2}}\mathds{1}_3\end{array}\right.$};
      \draw (0,-2) node [anchor=north] {$\theta=\frac{3\pi}{2}$, $\left\{\begin{array}{l}\Phi_\L=\tfrac{v}{\sqrt{2}}\mathds{1}_3\\\Phi_\R=-\sign(\gamma_1)\tfrac{v}{\sqrt{2}}\mathds{1}_3\end{array}\right.$};
      \draw [->, thick] (1.8,0) .. controls (1.5,2.3) and (-1.5,2.3)  .. (-1.8,0);
      \draw (0.9,0.9) node {kink};
    \end{tikzpicture}
    \caption{The kink soliton interpolates between the vacuum at
      $\theta=0$ and the sign-flipped vacuum at $\theta=\pi$, at the
      fine-tuned point in the theory: The two nonvanishing VEVs of
      $|\Phi_\L|=|\Phi_\R|$ equal each other up to a sign in this
      fine-tuned ground state.
      Like the kink soliton of fig.~\ref{fig:Josephson_vacua}, there
      is no solution interpolating between $\theta=0$ and
      $\theta=\frac\pi2$, but there is a solution going to
      $\Phi_\L=-\frac{v}{\sqrt{2}}\mathds{1}_3$,
      $\Phi_\R=-\sign(\gamma_1)\frac{v}{\sqrt{2}}\mathds{1}_3$, which
      however is not a ground state.
      The would-be ground states in the figure at $\theta=\frac\pi2$
      and $\frac{3\pi}{2}$ are thus marked with $\times$.
    }
    \label{fig:Josephson_fine-tuned_vacua}
  \end{center}
\end{figure}
We notice that both cosine terms in
eq.~\eqref{eq:eom_gamma1_theta_simpl} vanish if the coefficients are
fine tuned as $w_+=w_-$, which is a particularly fine-tuned point of
the parameters (vacuum) of the theory, see
fig.~\ref{fig:Josephson_fine-tuned_vacua}. 
This case is tantamount to
\beq
|\gamma_1|=\frac{m^2(\lambda_{34-1-2}+4\gamma_2)}{4\lambda_{12}},
\label{eq:gamma1_critical}
\eeq
for which the VEVs read
\beq
w_+=w_-=\frac{v}{\sqrt{2}} = \frac{m}{\sqrt{2\lambda_{12}}}.
\eeq
It also implies that the energies of the Josephson chirally
broken symmetric and asymmetric ground states become degenerate:
\beq
\calV =
-\frac{3m^4}{4\lambda_{12}}
-\frac{6\gamma_1^2}{\lambda_{34-1-2}+4\gamma_2}
=-\frac{3(m^2+2|\gamma_1|)^2}{2\xi}
=-\frac{3m^4\xi}{8\lambda_{12}^2}.
\eeq
The kink equation \eqref{eq:eom_gamma1_theta_simpl} now simplifies to
\beq
\delta(\d\theta)
+ |\gamma_1|\sin(2\theta)
- \frac12|\gamma_1|\sin(4\theta) = 0.
\eeq
This is the double sine-Gordon equation
\cite{Burt:1977ii,PhysRevB.27.474}, but with a negative coefficient of
the ``mass'' term.
The very particular combination of coefficients also makes it possible
to rewrite the equation as
\beq
\delta(\d\theta)
+2|\gamma_1|\sin^2(\theta)\sin(2\theta) = 0.
\eeq
Using a standard integration technique of multiplying the equation by
$\d\theta(\p_1)=\p_1\theta$ and integrating over $x$, we obtain the
quadratic form
\beq
(\p_1\theta)^2
= 2|\gamma_1|\sin^4\theta + A,
\label{eq:integrated_eom_fine_tuned}
\eeq
where we restrict the equation to the co-dimension one soliton case
and $A$ is an integration constant.
Setting $A:=0$ and taking the square root of the above equation yields
a simple differential equation with the solution
\beq
\theta(x) = \arctan\left(\pm\sqrt{2|\gamma_1|}(x-x_0)\right)+\frac\pi2.
\label{eq:kink_sol_fine_tuned}
\eeq
This exact analytic solution at the special fine-tuned point of the
theory, is not the standard sine-Gordon solution, but instead simply a
simple (inverse) trigonometric function.
This solution should be contained in the double-sine Gordon solutions
as well.

The energy (tension) of this fine-tuned kink is given by
\beq
E = 3\pi v^2\sqrt{2|\gamma_1|},
\eeq
where we have set the ground state energy to zero.

Finally, we note that both the equation
\eqref{eq:integrated_eom_fine_tuned} and the solution
\eqref{eq:kink_sol_fine_tuned} demonstrate that this fine-tuned point
in the theory contains again a ``kink''-type solution: That is, the
solution interpolates between the ground state and the parity-flipped
ground state (which is physically equivalent to the ground state), but
not the other would-be ground state, see $\times$ in 
fig.~\ref{fig:Josephson_fine-tuned_vacua}.

\section{Numerical solutions of domain walls}\label{sec:numerical}

The sigma-model limit studied above is very powerful as it fixes the
complex scalar fields to be unitary matrices\footnote{In the vacuum
with $\gamma_1\neq0$, this is not necessarily true, see the discussion
below. } and determines the gauge
field explicitly.
This and the appropriate usage of gauge and flavor symmetries, reduces
the entire problem to a sine-Gordon equation when the $\gamma$ terms
are turned off.
Turning on $\gamma_3\neq0$ yields a modified sine-Gordon equation.
Turning on the Josephson ($\gamma_1$) term complicates matters; indeed
we have shown in theorem \ref{thm:1} that there are no DW solutions
that swap the vacua of $\Phi_\L\leftrightarrow\Phi_\R$, but there is a
``kink'' solution that goes over the lifted vacuum and returns to the
same (parity flipped) ground state.
We have also found an exact analytic solution for a fine-tuned point
in the theory, which is again of the ``kink'' type.

Without taking the sigma-model limit, but taking generic values of the
couplings in the theory -- gauge coupling and potential couplings --
we need to perform numerical computations for obtaining the DWs or
kinks; that is, by solving the full equations of motion
\eqref{eq:eom_PhiL}-\eqref{eq:eom_A}.
We will compare the solutions of the full equations of motion to the
sigma-model solutions obtained in the previous section (analytic or
numeric, depending on the parameters), to see when and how good
approximations to the true solutions they are.

First we will establish the following lemma.
\begin{lemma}\label{lem2}
  Real-valued diagonal solutions to the complex scalar fields $\Phi_\L$ and
  $\Phi_\R$ imply vanishing gauge fields, $A=0$.
\end{lemma}
\emph{Proof}:
Real-valued diagonal fields satisfy $\Phi_\L=\Phi_\L^\dag$ and
$\Phi_\R=\Phi_\R^\dag$, which reduce the equation of motion for the
gauge field \eqref{eq:eomA} to
\beq
\Tr\left[\{A^a t^a, \Phi_\L\Phi_\L^\dag + \Phi_\R\Phi_\R^\dag\}t^b\right]t^b = 0,
\eeq
which for diagonal $\Phi_\L\Phi_\L^\dag+\Phi_\R\Phi_\R^\dag$ only has
the solution $A=A^at^a=0$, where we have used the Gell-Mann matrices
as an $\SU(3)$ basis with $\Tr[t^a t^b]=-2\delta^{ab}$,
$a,b=1,\ldots,8$ and $t^a=-(t^a)^\dag$. 
\hfill$\square$

In ref.~\cite{Gudnason:2025qxf}, Lemma 1, we proved that if the scalar
and gauge fields are diagonal, they will remain diagonal under
gradient flow or Newton-flow evolution of the equations of motion
($A=0$ is a special case of a diagonal matrix).

It is easy to observe by inspection of the equations of motion for the
scalar fields, $\Phi_\L$ and $\Phi_\R$, that if the fields are
real-valued, they will remain real-valued under gradient flow or
Newton flow, if and only if the gauge field vanishes: $A=0$ (we recall
that all couplings $m$, $\lambda_{1,2,3,4}$ and $\gamma_{1,2,3}$ are
real-valued).

Using this fact, as well as Lemma \ref{lem2} and Lemma 1 of
ref.~\cite{Gudnason:2025qxf}, we can write the following corollary.
\begin{corollary}
If the boundary conditions allow for a diagonal and real-valued
initial guess for the scalar fields $\Phi_\L$ and $\Phi_\R$ with a
vanishing initial gauge field $A=0$, then the configuration will flow
to a solution to the equations of motion that is also diagonal and
real valued.
\end{corollary}
This corollary helps us verify that most of the soliton solutions will
in fact have a vanishing gauge field.

Interestingly, the solution \eqref{eq:Asol_general} is nonvanishing
for nonconstant $\Phi_{\L,\R}$, which is true under the assumption of
the sigma-model limit\footnote{We should note that a constant $W$ of eq.~\eqref{eq:DWansatz} yields again a vanishing gauge field: $A=0$, see eq.~\eqref{eq:Asol_sigma_model_limit}. }.
Of course, the sigma-model limit is only an approximation facilitating
analytic computations, but as we will see shortly it is indeed a quite
useful simplification catching certain main aspects of the true
solutions.

\subsection{Domain walls between chirally symmetric ground states}

We start with the case of the DW solitons in the theory with the
$\gamma$-part of the potential turned off
($\gamma_1=\gamma_2=\gamma_3=0$).
The DW thus interpolates between two different ground states studied in
sec.~\ref{sec:lvac}.

\begin{figure}[!htp]
  \centering
  \mbox{\subfloat[]{\includegraphics[width=0.49\linewidth]{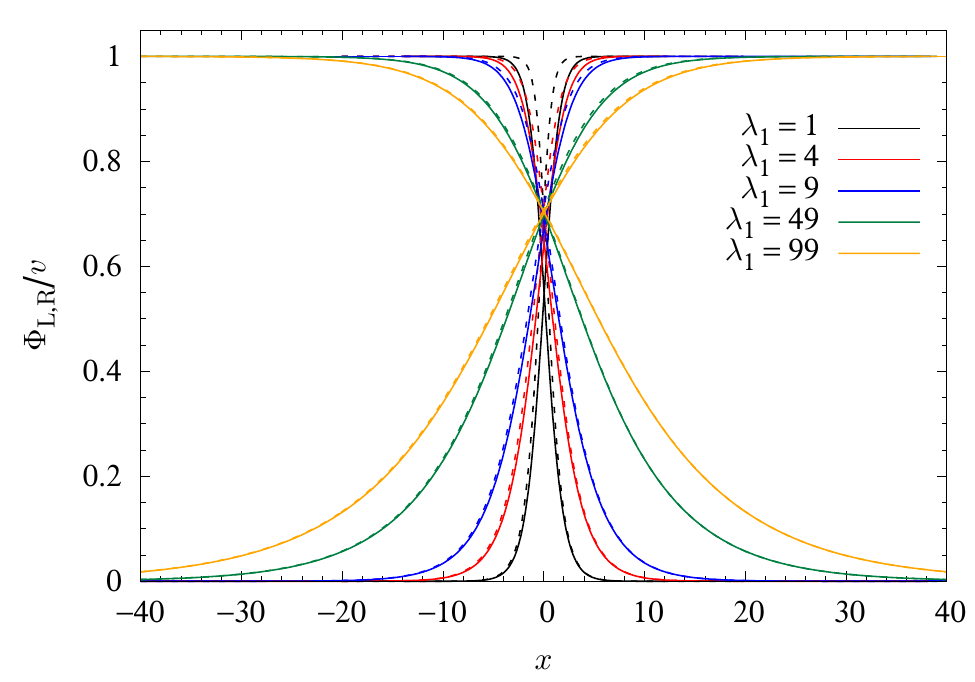}}
    \subfloat[]{\includegraphics[width=0.49\linewidth]{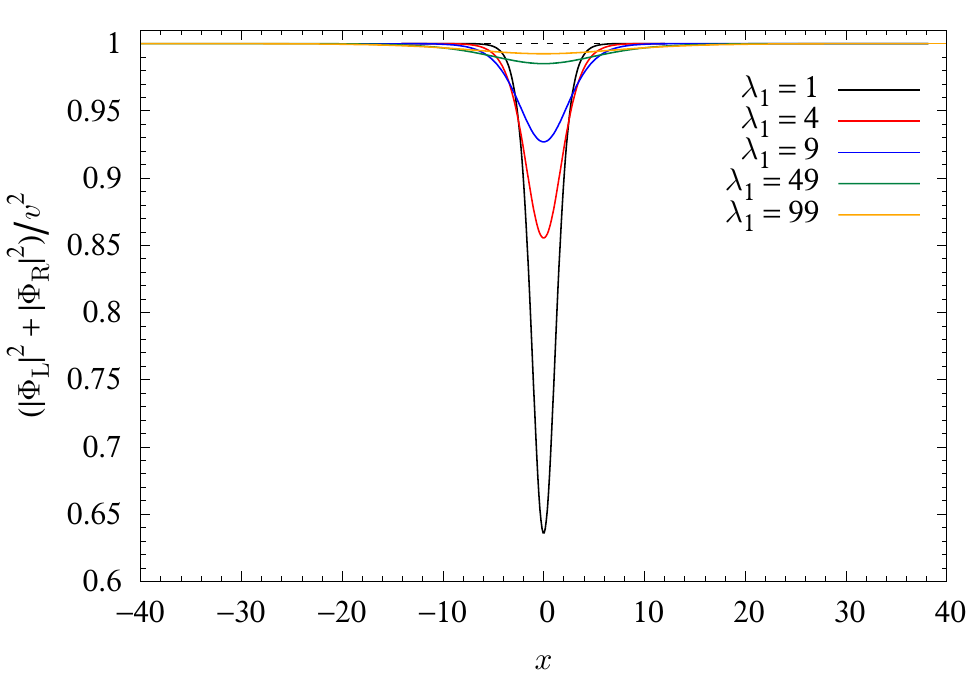}}}
    \mbox{\sidesubfloat[]{\includegraphics[width=0.49\linewidth]{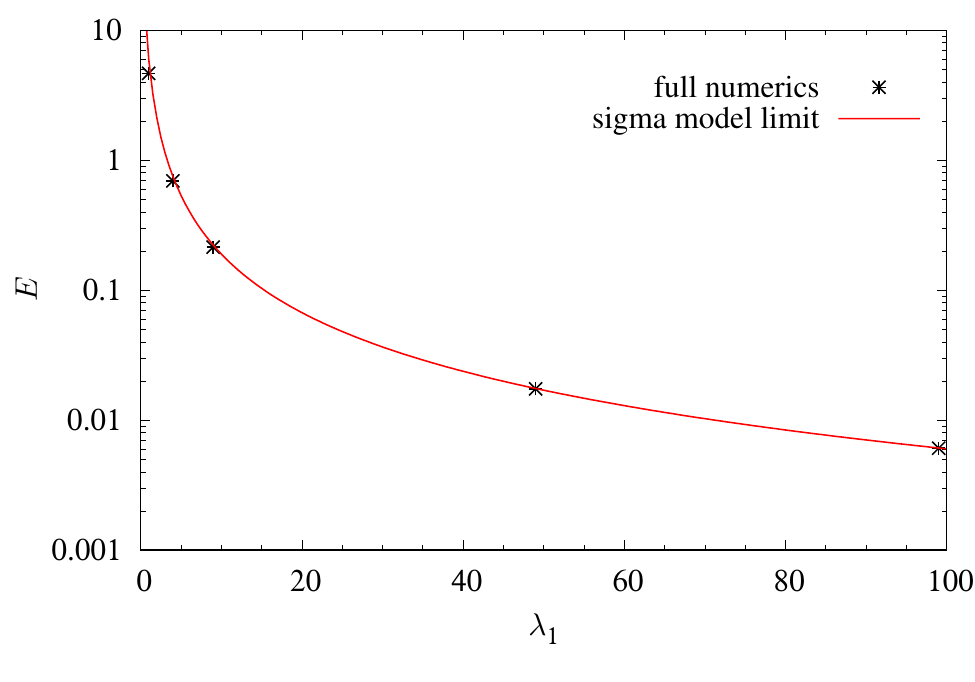}}}
  \caption{DWs in the chirally symmetric but asymmetric ground states
    interpolating between $(\Phi_\L,\Phi_\R)=(v\mathds{1}_3,0)$ and
    $(\Phi_\L,\Phi_\R)=(0,v\mathds{1}_3)$ (see
    fig.~\ref{fig:DW_vacua}), for a variety of couplings 
    $\lambda_1=1,4,9,49,99$: (a) the diagonal part (all three elements
    are equal) of the scalar fields $\Phi_{\L,\R}$, (b) the sigma-model
    constraint \eqref{eq:sigma_model_constraint}.
  We compare the full computations (solid lines) with
  the sigma-model limit (dashed lines) and take
  $\lambda_4=\lambda_1+1$ so that $\lambda_4-\lambda_1=1>0$, but
  $\lambda_1$ is increased (the sigma-model limit corresponds to
  $\lambda_1\to\infty$).
   (c) Total energy (tension) of the DW with the solid line displaying the sigma-model limit result \eqref{eq:Esigmamodel_lambda} and the points showing the energies of full computations.
  In this figure $m=\sqrt{2}$, $\lambda_{2,3}=0$,
  $\lambda_4=\lambda_1+1$, $\gamma_{1,2,3}=0$. 
  }
  \label{fig:lvac}
\end{figure}
In fig.~\ref{fig:lvac}(a,b) full numerical computations are shown with solid
lines that should be compared with the sigma-model limit result
\eqref{eq:SG_lambda_sol}, shown with dashed lines.
In order to take the sigma-model limit \eqref{eq:sigma_model_limit}, we increase the value of $\lambda_1$ and keep $\lambda_4-\lambda_1>0$ fixed (and positive); more precisely, we fix here $\lambda_4-\lambda_1=1$.
Fig.~\ref{fig:lvac}(a) displays the scalar fields and one can observe
that for the larger values of the coupling $\lambda_1$, the solid and
dashed lines converge approximately.
Fig.~\ref{fig:lvac}(b) displays the sigma-model constraint
\eqref{eq:sigma_model_constraint}, i.e.~if the lines equal unity the
sigma-model constraint is satisfied.
Clearly, the constraint is not satisfied at all for $\lambda_1=1$,
but the approximation becomes better and better for larger values of
the coupling.
Finally, in fig.~\ref{fig:lvac}(c) we show the DW energy (tension) of the sigma-model limit \eqref{eq:Esigmamodel_lambda} with a solid line and the energies of the full computations with points.
The energies are also converging to the sigma-model limit result from below for large $\lambda_1$ (recall that we have fixed $\lambda_4-\lambda_1=1$ here). 
This means that the sigma-model limit overestimates the DW energy (tension) for small values of $\lambda_1$.

\subsection{Domain walls in Josephson-free chirally broken ground states \texorpdfstring{($\gamma_1=0$, $\gamma_2\leq0$, $\gamma_3\neq0$)}{(gamma1=0, gamma2<=0, gamma3=/=0)}}

We will now turn to the determinant or $\gamma_3$ term in the
potential \eqref{eq:V}, which leaves the ground states unchanged with
respect to the previous subsection, but changes the condition for the
ground states to be true ground states and hence the condition for the
existence of the DW, see sec.~\ref{sec:vac_gamma23}.
We remark that the ground states remain simple only for
$\gamma_2\leq0$; for $\gamma_2>0$ complicated complex phases are
introduced and the vacuum structure complicates the problem
significantly.

\begin{figure}[!htp]
  \centering
  \mbox{\subfloat[]{\includegraphics[width=0.49\linewidth]{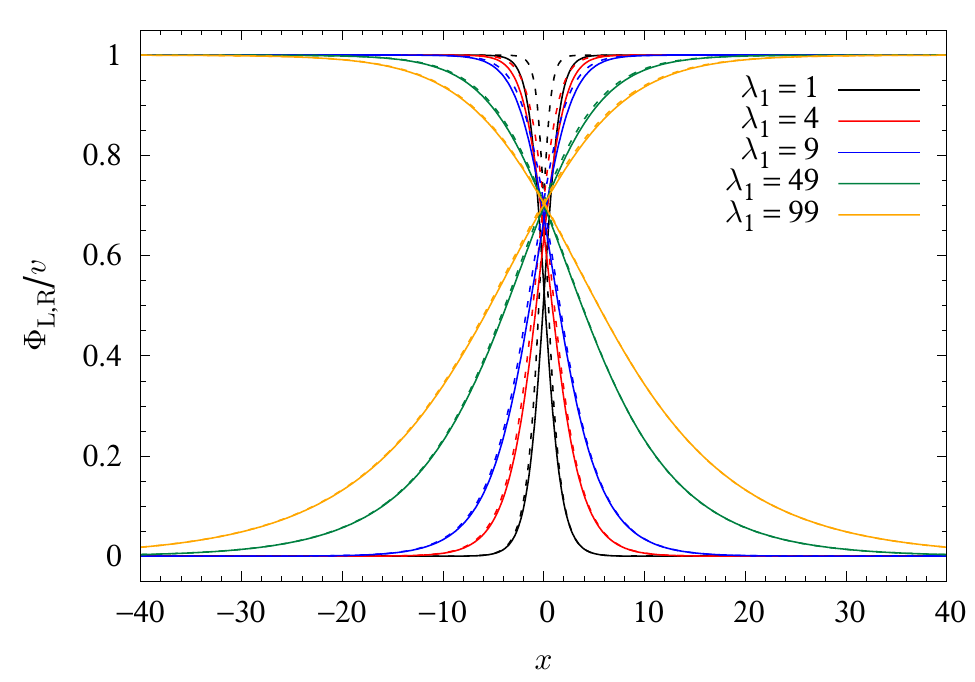}}
    \subfloat[]{\includegraphics[width=0.49\linewidth]{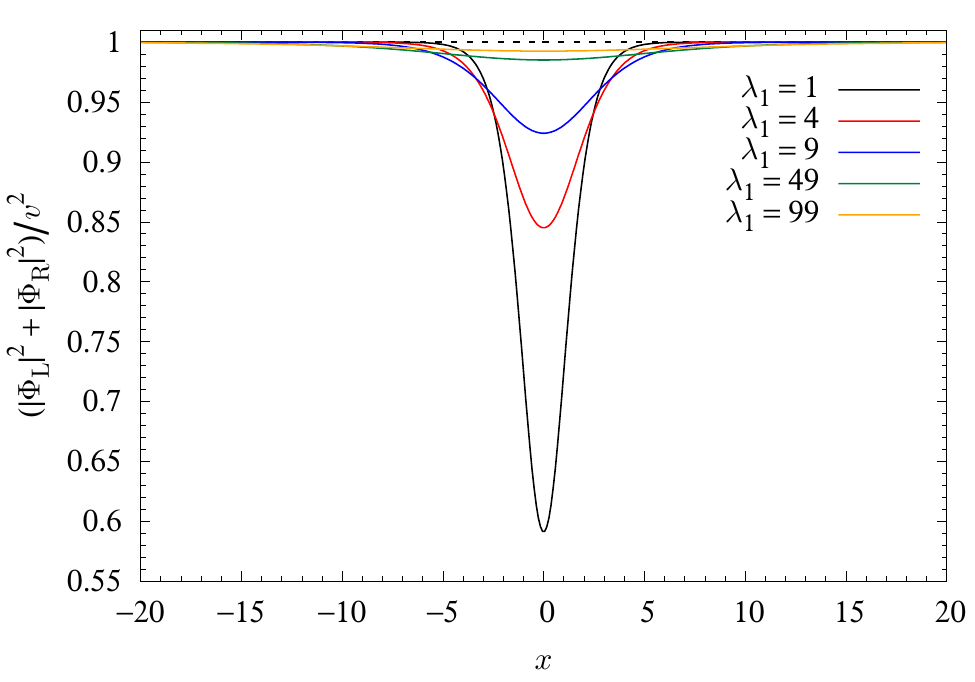}}}
  \mbox{\subfloat[]{\includegraphics[width=0.49\linewidth]{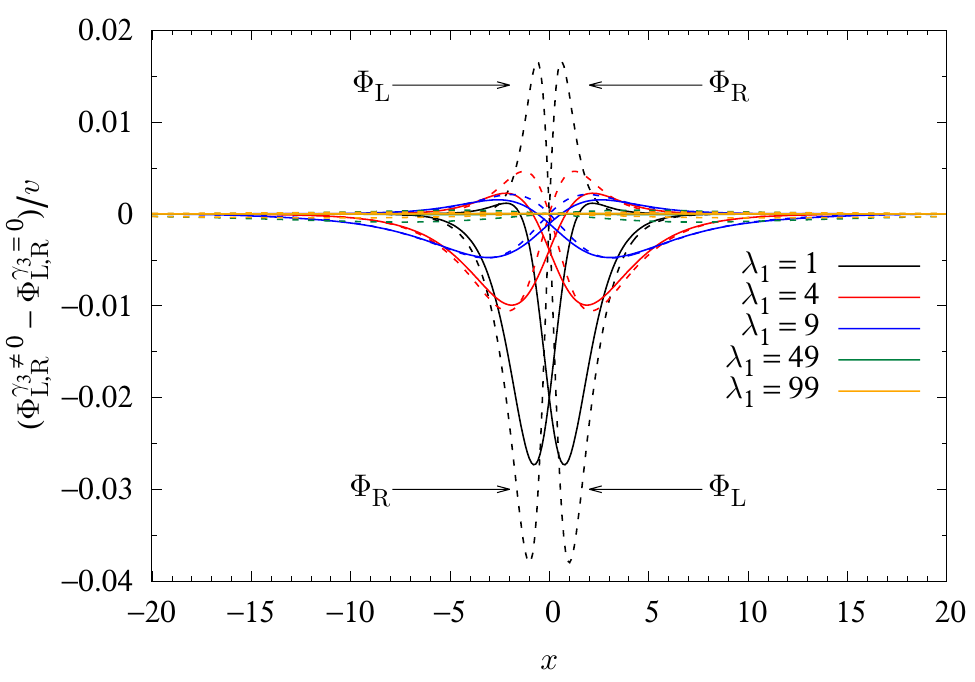}}
    \subfloat[]{\includegraphics[width=0.49\linewidth]{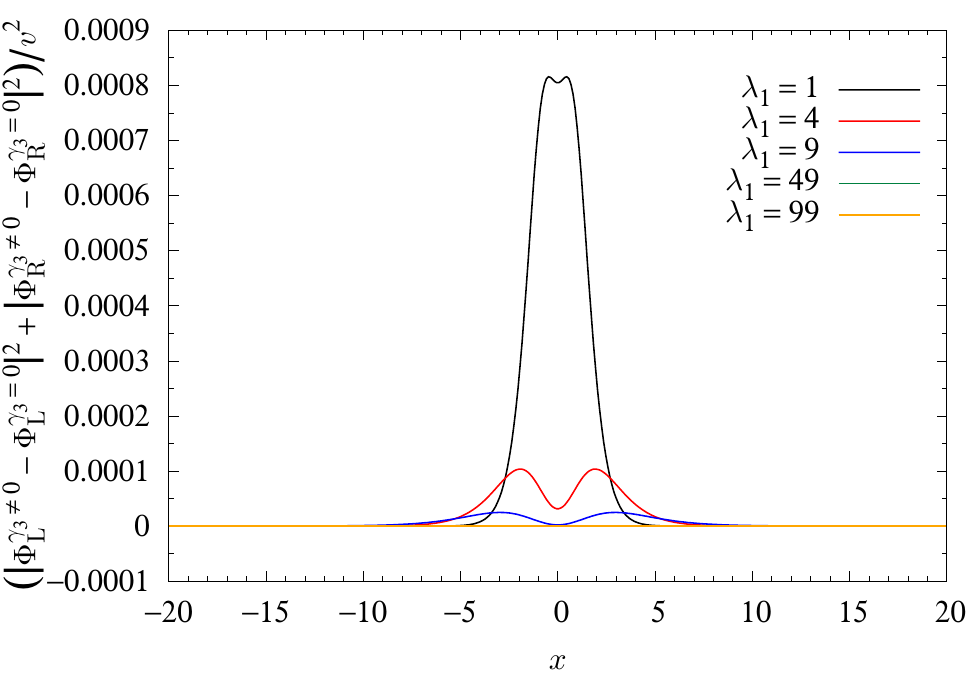}}}
  \mbox{\subfloat[]{\includegraphics[width=0.49\linewidth]{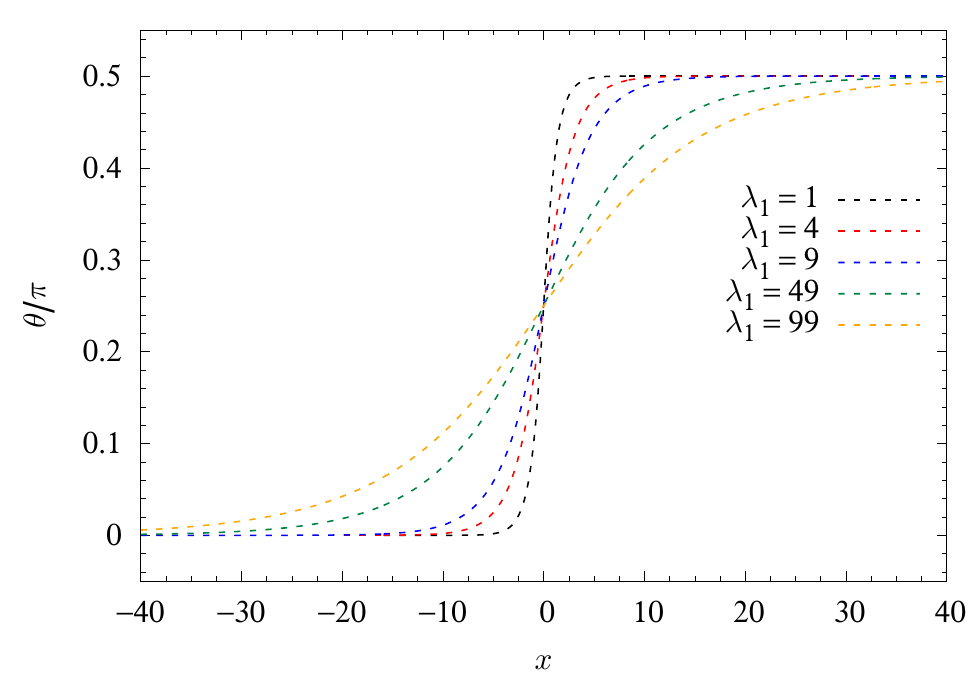}}
  \subfloat[]{\includegraphics[width=0.49\linewidth]{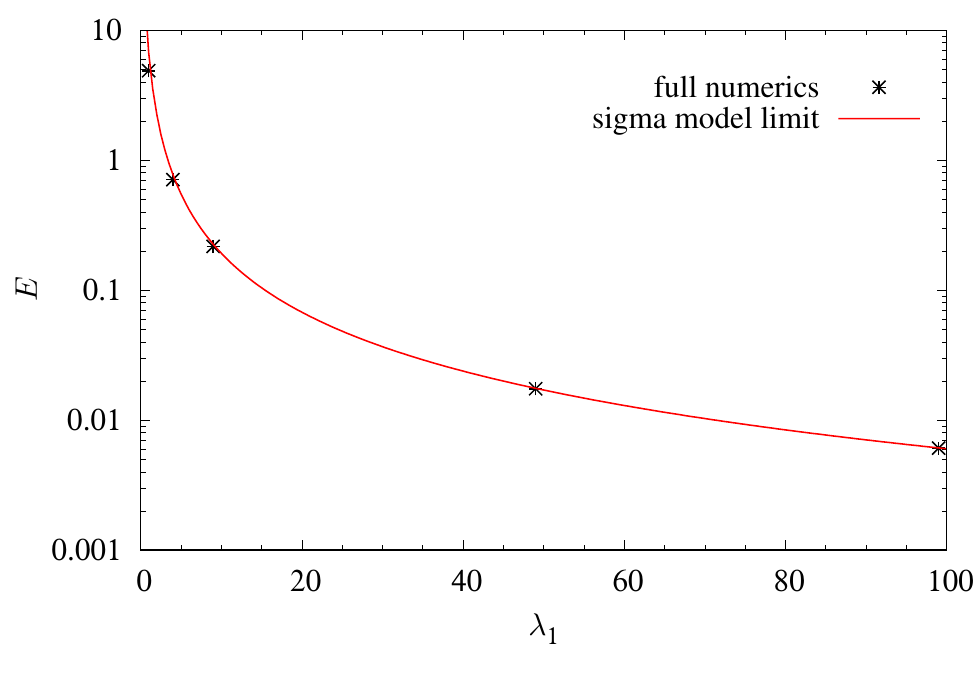}}}
  \caption{DWs in the chirally broken and asymmetric ground states
    interpolating between $(\Phi_\L,\Phi_\R)=(v\mathds{1}_3,0)$ and
    $(\Phi_\L,\Phi_\R)=(0,v\mathds{1}_3)$, for a variety of couplings 
    $\lambda_1=1,4,9,49,99$: (a) the diagonal part (all three elements
    are equal) of the scalar fields $\Phi_{\L,\R}$, (b) the sigma-model
    constraint \eqref{eq:sigma_model_constraint}, (c) the scalar
    fields with $\gamma_3=\tfrac14$ minus the corresponding
    $\gamma_3=0$ solution
    ($\Phi_{\L,\R}^{\gamma_3=1/4}-\Phi_{\L,\R}^{\gamma_3=0}$),
    (d) the sigma model constraint \eqref{eq:sigma_model_constraint}
    computed for the difference of fields displayed in panel (c).
  We compare the full computations (solid lines) with
  the sigma-model limit (dashed lines).
  (e) The sigma-model limit result $\theta$.
  (f) Total energy (tension) of the DW with the solid line displaying the sigma-model limit and the points showing the energies of full computations.
  In this figure $m=\sqrt{2}$, $\lambda_{2,3}=0$,
  $\lambda_4=\lambda_1+1$, $\gamma_{1,2}=0$ and $\gamma_3=\tfrac14$.
  }
  \label{fig:g23vac}
\end{figure}
In fig.~\ref{fig:g23vac}(a-d) full numerical computations are shown with
solid lines compared with the numerical solution in the sigma-model limit (shown with dashed lines),
which is a solution to eq.~\eqref{eq:eom_gamma_theta_simpl}.
In this example, we have fixed $\gamma_3=\tfrac14$ and taken the same
values of $\lambda_{1,4}$ as in fig.~\ref{fig:lvac}; again setting
$\lambda_4=\lambda_1+1$. 
Fig.~\ref{fig:g23vac}(a) displays the scalar fields and for the larger
values of the coupling $\lambda_1$, the solid and dashed lines
converge approximately also in this case.
Fig.~\ref{fig:g23vac}(b) displays the sigma-model constraint
\eqref{eq:sigma_model_constraint}, i.e.~if the lines equal unity the
sigma-model constraint is satisfied.
Since the numerical solutions -- both the full computations and the
sigma-model limit solutions -- look almost identical to those of
fig.~\ref{fig:lvac}, we display the difference between the two sets of
solutions in panels (c) and (d).
That is, in fig.~\ref{fig:g23vac}(c) is shown the difference between
the solutions with $\gamma_3=\tfrac14$ and those with $\gamma_3=0$ and 
in fig.~\ref{fig:g23vac}(d) is shown the sigma-model
constraint \eqref{eq:sigma_model_constraint} of the same difference
between solutions.
Since the sigma-model limit solution is no longer just the sine-Gordon solution, we also show the profile $\theta$ in fig.~\ref{fig:g23vac}(e), which is a solution to eq.~\eqref{eq:eom_gamma_theta_simpl}.
Finally, we compute the DW energy (tension) of the sigma-model limit solution and the full numerical computation and display them in fig.~\ref{fig:g23vac}(f) with a solid red line and points, respectively.
Since the difference between the $\gamma_3=\tfrac14$ solutions and the
$\gamma_3=0$ solutions is very small, indeed the convergence
properties in the limit of large $\lambda_1$ with
$\lambda_4-\lambda_1=1$ fixed, are equally good.

\begin{figure}[!htp]
  \centering
  \mbox{\subfloat[]{\includegraphics[width=0.49\linewidth]{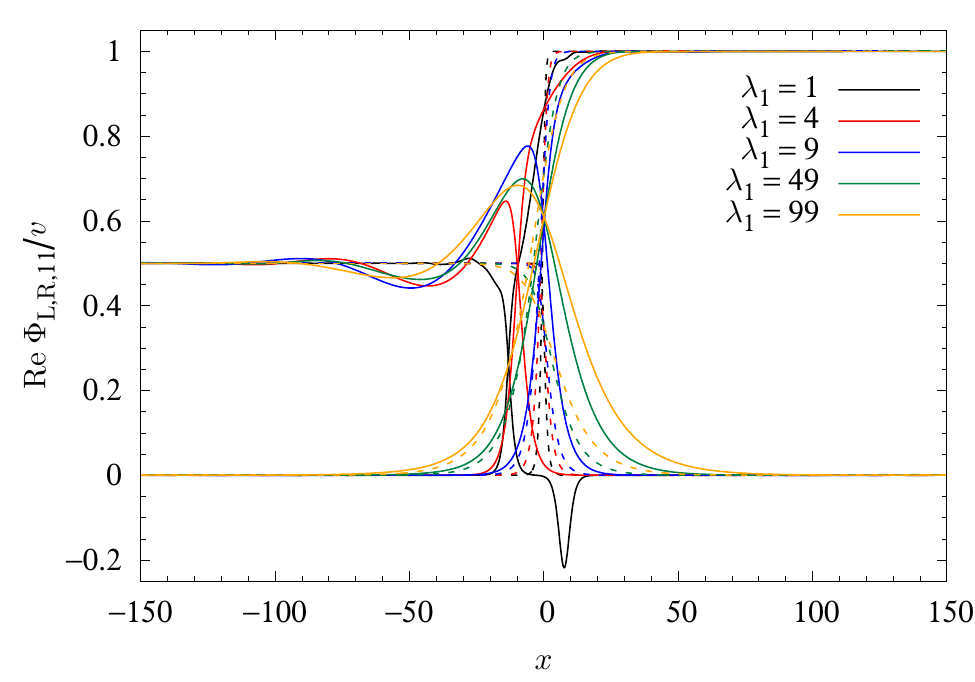}}
    \subfloat[]{\includegraphics[width=0.49\linewidth]{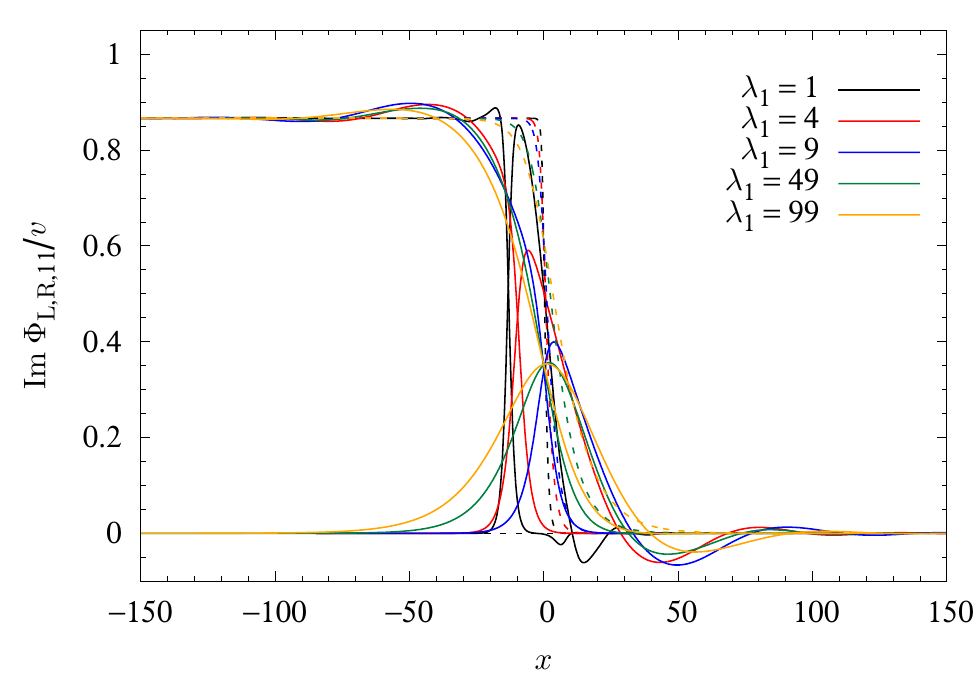}}}
  \mbox{\subfloat[]{\includegraphics[width=0.49\linewidth]{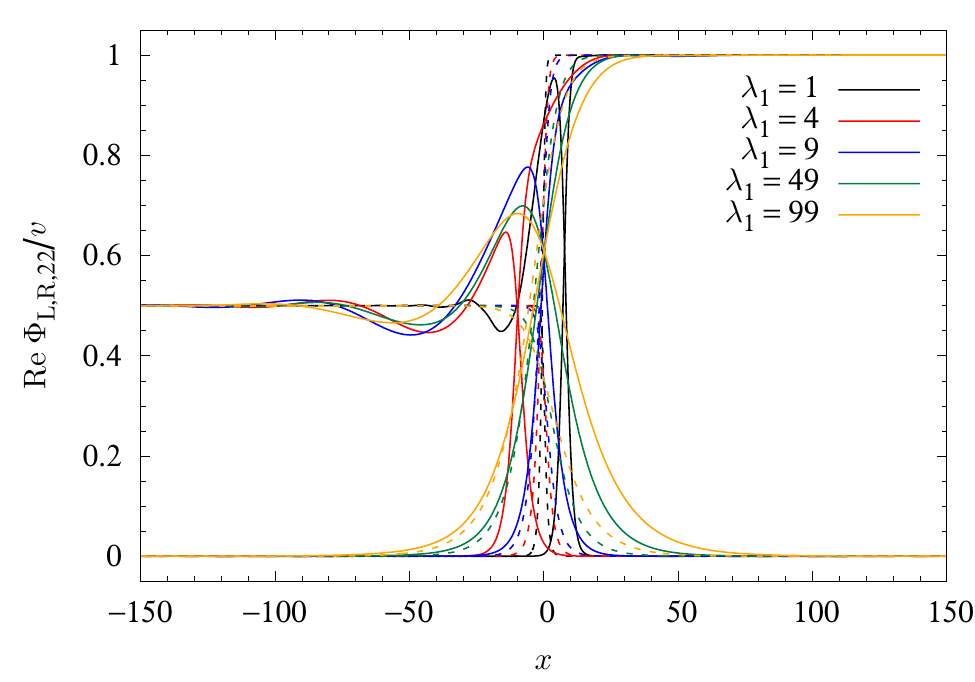}}
    \subfloat[]{\includegraphics[width=0.49\linewidth]{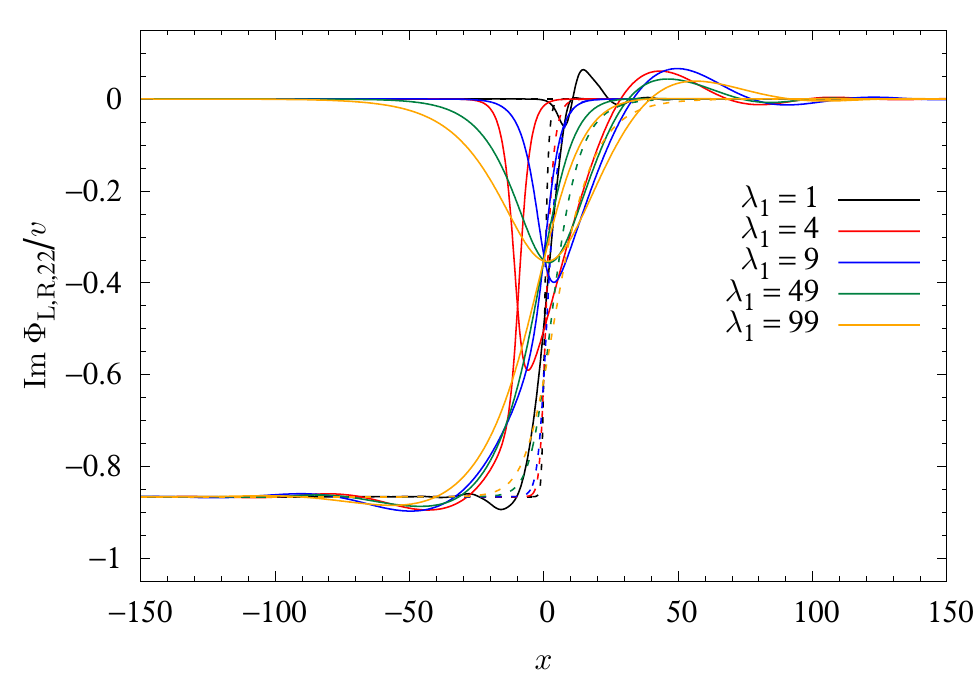}}}
  \mbox{\sidesubfloat[]{\includegraphics[width=0.49\linewidth]{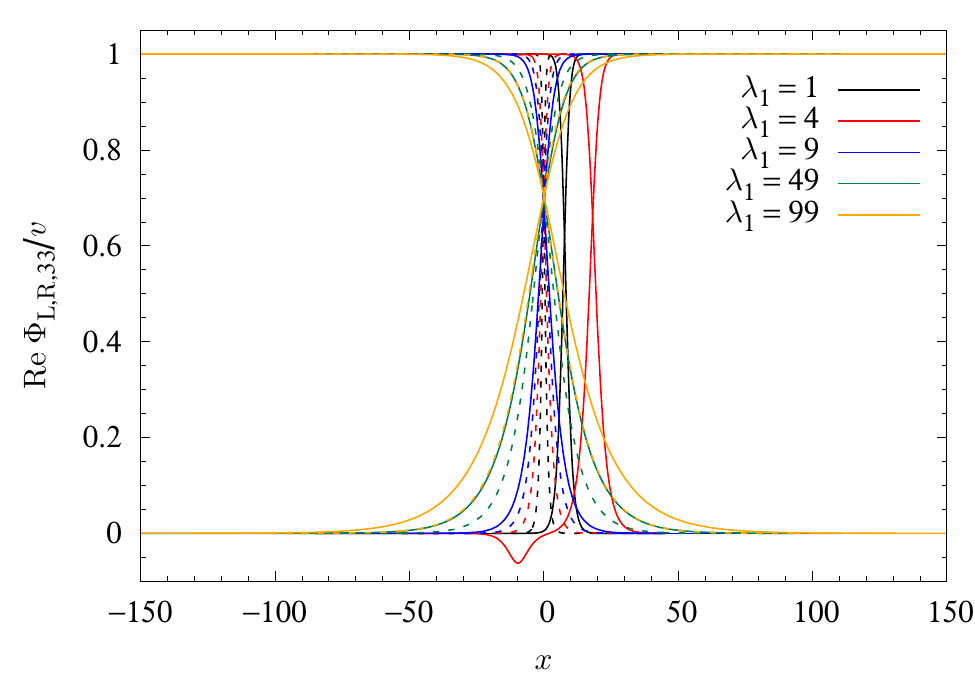}}}
  \caption{DWs in the chirally broken and asymmetric ground states
    interpolating between $(\Phi_\L,\Phi_\R)=(v\diag(e^{\i\pi/3},e^{-\i\pi/3},1),0)$ and
    $(\Phi_\L,\Phi_\R)=(0,v\mathds{1}_3)$, for a variety of couplings 
    $\lambda_1=1,4,9,49,99$.
    (a) and (b) show the real and imaginary parts of the 11-component of $\Phi_{\L,\R}$. (c) and (d) show the real and imaginary part of the 22-component of $\Phi_{\L,\R}$. (e) shows the real part of the 33-component of $\Phi_{\L,\R}$ while the imaginary part vanishes.
    The sigma-model limit is shown with dashed lines. 
    In this figure $m=\sqrt{2}$, $\lambda_{2,3}=0$,
  $\lambda_4=\lambda_1+1$, $\gamma_{1}=0$, $\gamma_2=-\tfrac18$ and $\gamma_3=\tfrac14$.}
  \label{fig:g23vac_alpha_phi}
\end{figure}

\begin{figure}[!htp]
  \centering
  \mbox{\subfloat[]{\includegraphics[width=0.49\linewidth]{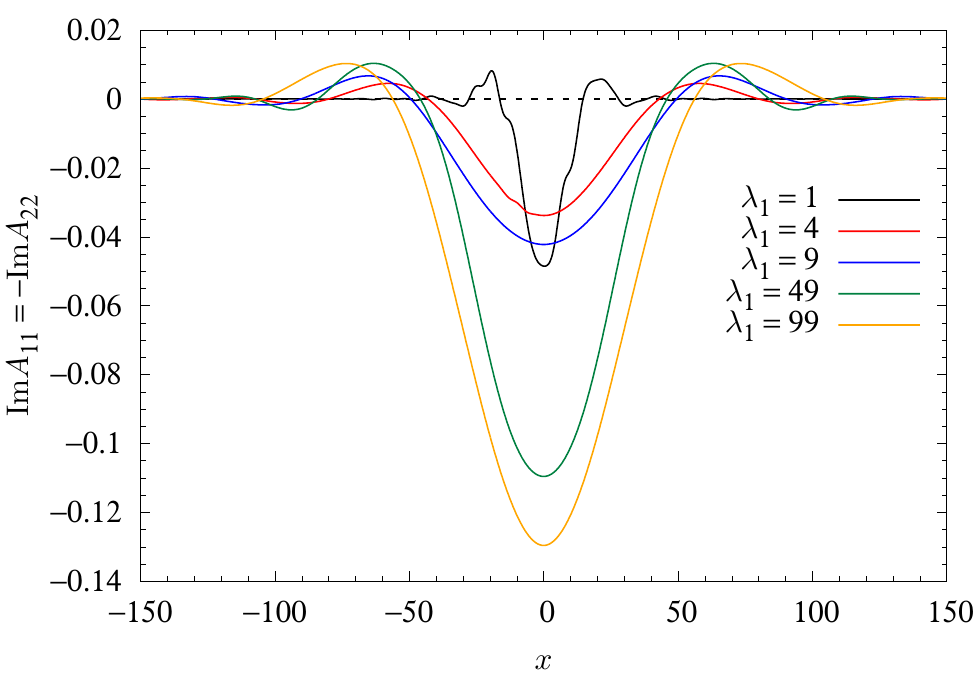}}
    \subfloat[]{\includegraphics[width=0.49\linewidth]{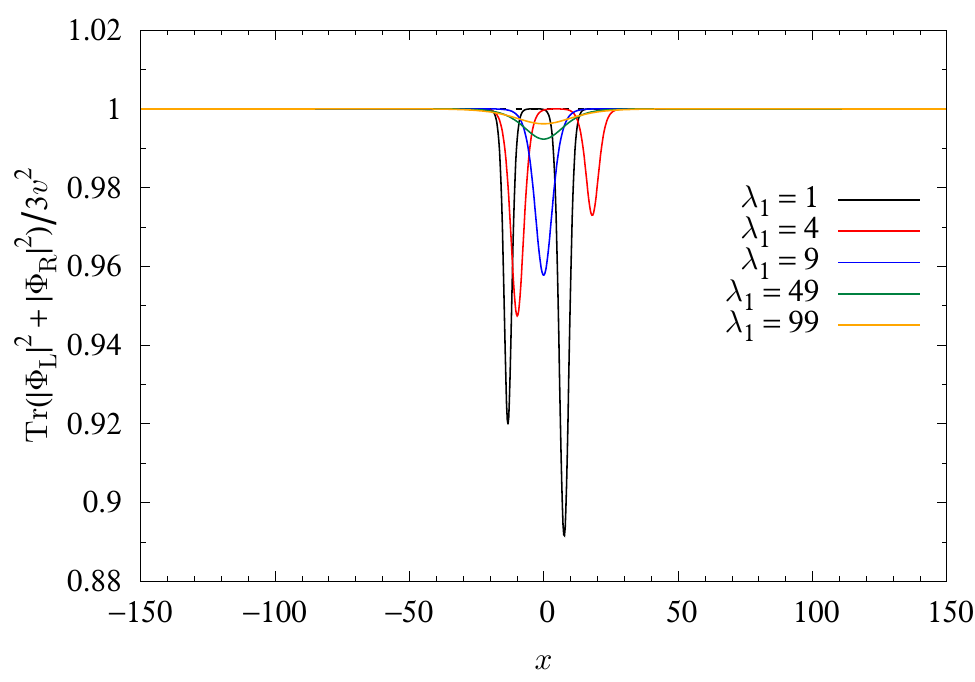}}}
  \mbox{\sidesubfloat[]{\includegraphics[width=0.49\linewidth]{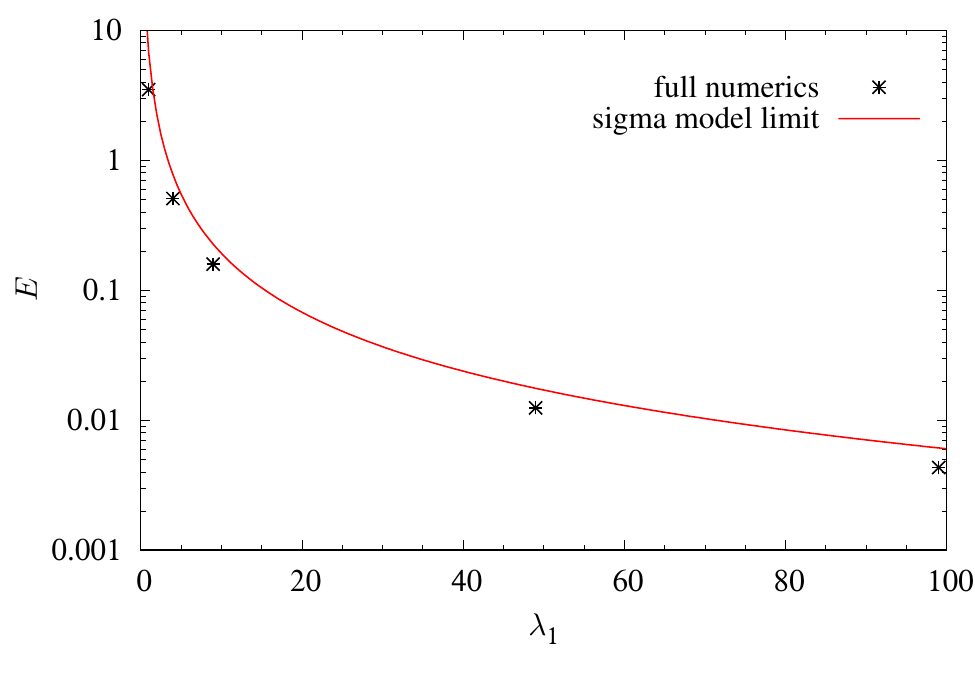}}}
  \caption{DWs in the chirally broken and asymmetric ground states
    interpolating between $(\Phi_\L,\Phi_\R)=\big(v\diag(e^{\i\pi/3},e^{-\i\pi/3},1),0\big)$ and
    $(\Phi_\L,\Phi_\R)=(0,v\mathds{1}_3)$, for a variety of couplings 
    $\lambda_1=1,4,9,49,99$.
    (a) the 11-component of the gauge field (which is equal to minus the 22-component).
    (b) The sigma model constraint \eqref{eq:sigma_model_constraint}.
    (c) The energy of the sigma-model limit (solid line) compared with the full numerical computations (points). 
    In this figure $m=\sqrt{2}$, $\lambda_{2,3}=0$,
  $\lambda_4=\lambda_1+1$, $\gamma_{1}=0$, $\gamma_2=-\tfrac18$ and $\gamma_3=\tfrac14$.}
  \label{fig:g23vac_alpha_A_norm_energy}
\end{figure}

In the case with both the $\gamma_3$ and $\gamma_2<0$ terms turned on, it is possible to alter the boundary conditions of the field $W$, such that it is not just the identity matrix, but contains complex phases (but still with $\det W=1$); this corresponds to considering $\alpha\neq0$ of eq.~\eqref{eq:Wsols} (if $\gamma_2=0$, it has no effect on the solutions, as $\det W=1$ always).
We will here consider the special case of $\alpha$ being a solution to $1+2\cos(2\alpha)=0$, which makes the sigma-model limit unaware of the $\gamma_2$ term -- it has exactly the same sigma-model limit, as shown in fig.~\ref{fig:g23vac}(e). 
The full field theory computation turns out to be more complicated and the true solutions are not captured well by the sigma-model limit in this case, see figs.~\ref{fig:g23vac_alpha_phi} and \ref{fig:g23vac_alpha_A_norm_energy}.
In fig.~\ref{fig:g23vac_alpha_phi} are shown the real and imaginary parts of the diagonal (11, 22 and 33) components of the scalar fields $\Phi_{\L,\R}$, except the imaginary part of the 33-component, as it vanishes.
The gauge field is now turned on, since there are nontrivial complex phases of the scalar fields in play. Since the boundary condition with $\alpha\neq0$ is chosen only to affect the first two diagonal components of the scalar field $\Phi_\L$, the 33-component of the gauge field remains vanishing. Since the gauge field is imaginary and traceless, the two first diagonal components are equal but with opposite signs, see fig.~\ref{fig:g23vac_alpha_A_norm_energy}(a).
Although this example does not show good convergence to the sigma-model limit's prediction, the sigma model constraint approaches unity and hence is expected to be obeyed in the limit of $\lambda_1\to\infty$.
As the convergence is poor, also the DW energy (tension) does not converge well to the sigma-model limit, see fig.~\ref{fig:g23vac_alpha_A_norm_energy}(c).
The assumptions of the different components being relatively constant, i.e.~that $W$ remains constant, is not a good approximation for finite values of the coupling, $\lambda_1$.

\subsection{Josephson chirally broken ground states
  \texorpdfstring{($\gamma_1\neq0$, $\gamma_2\in\mathbb{R}$, $\gamma_3=0$)}{(gamma1=/=0, gamma2 in R, gamma3=0)}: Kink}

We now turn to the case of the ground states with nonvanishing
Josephson coupling, $\gamma_1\neq0$.
In this case, there is no DW that interpolates between the asymmetric
ground state and the flavor-swapped ground state
$\Phi_\L\leftrightarrow\Phi_\R$, see fig.~\ref{fig:Josephson_vacua}.
There is, in principle, a soliton that goes to a swapped state with
the opposite relative sign between the left and right scalar fields;
such state is however not a ground state; it is thus marked with a
$\times$ in fig.~\ref{fig:Josephson_vacua}.
Nevertheless, there is a soliton solution that interpolates between
the asymmetric ground state and its overall sign-flipped version, which
we denote a kink soliton.

\begin{figure}[!htp]
  \centering
  \mbox{\subfloat[]{\includegraphics[width=0.49\linewidth]{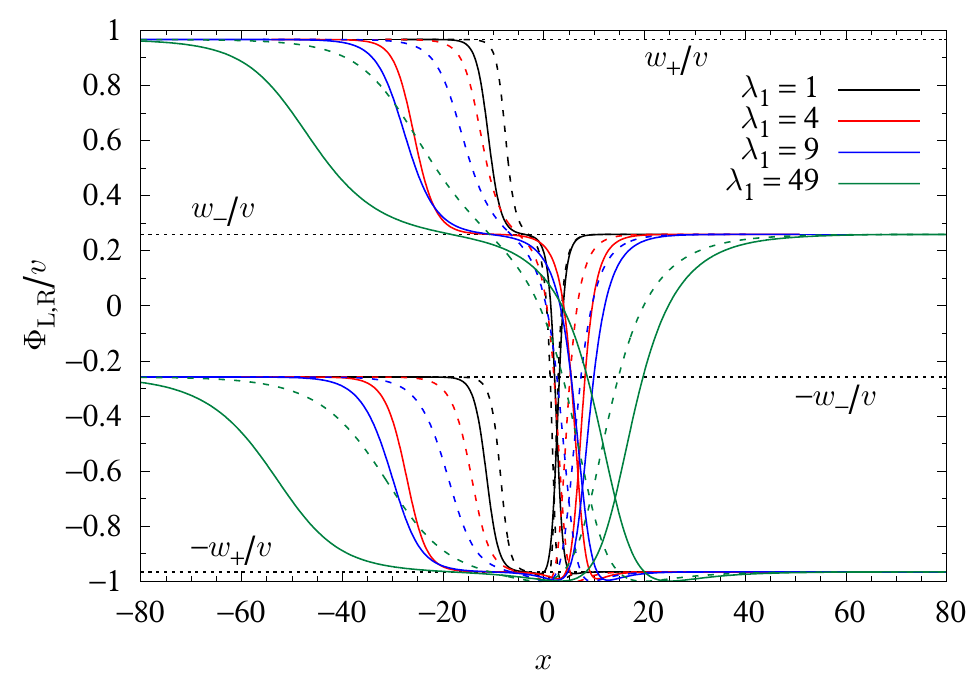}}
    \subfloat[]{\includegraphics[width=0.49\linewidth]{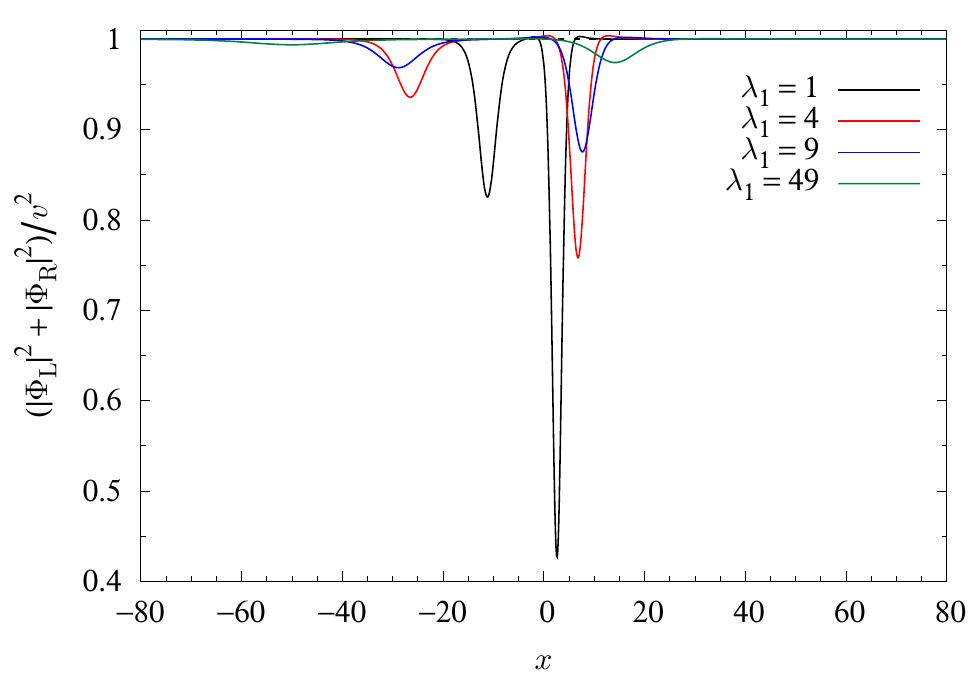}}}
  \mbox{\subfloat[]{\includegraphics[width=0.49\linewidth]{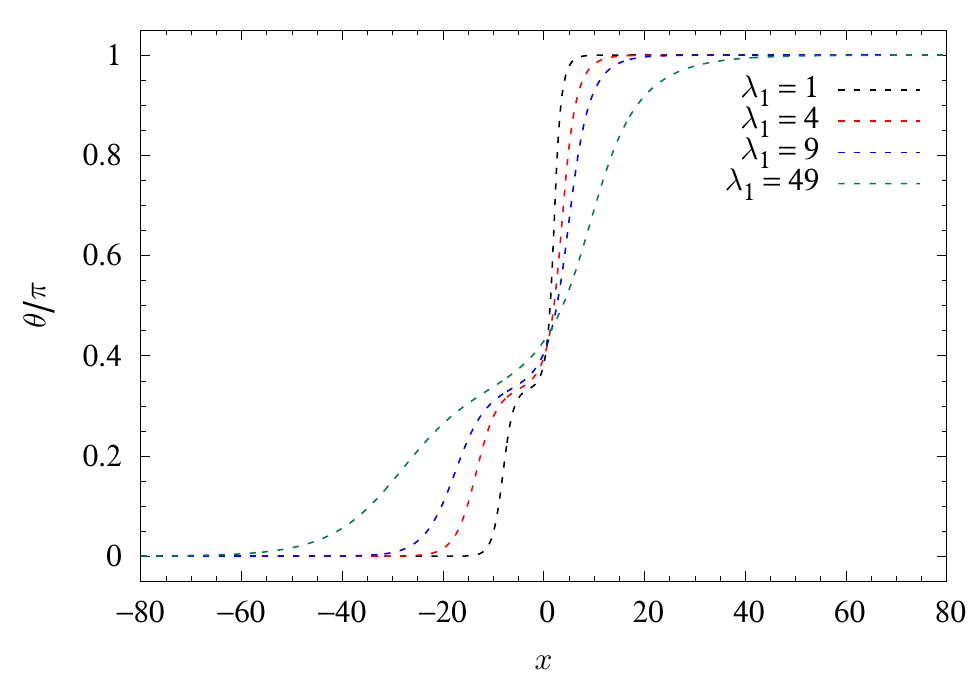}}
  \subfloat[]{\includegraphics[width=0.49\linewidth]{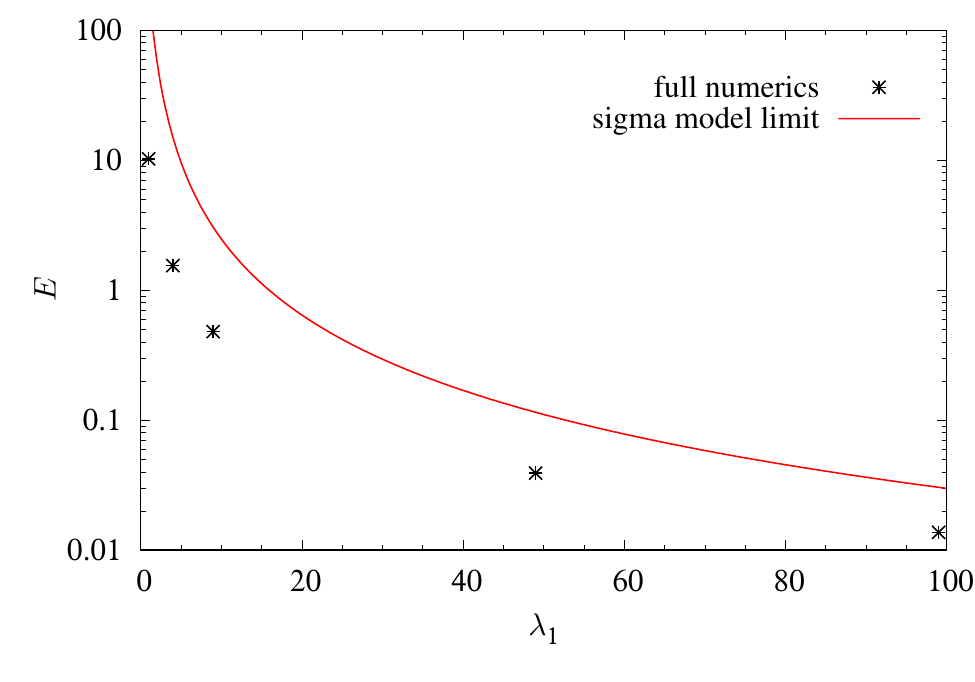}}}
  \caption{Kink solitons in the chirally broken Josephson ground states
    interpolating between
    $(\Phi_\L,\Phi_\R)=(w_+\mathds{1}_3,-\sign(\gamma_1)w_-\mathds{1}_3)$
    and 
    $(\Phi_\L,\Phi_\R)=(-w_+\mathds{1}_3,\sign(\gamma_1)w_-\mathds{1}_3)$
    (see fig.~\ref{fig:Josephson_vacua}), for a variety of couplings 
    $\lambda_1=1,4,9,49,99$: (a) the diagonal part (all three elements
    are equal) of the scalar fields $\Phi_{\L,\R}$, (b) the sigma-model
    constraint \eqref{eq:sigma_model_constraint} and (c) the
    sigma-model limit's soliton profile function $\theta$,
    interpolating between $0$ and $\pi$.
    We compare the full computations (solid lines) with
    the sigma-model limit (dashed lines) [except in panel (c), where
      the full computations do not have a well-defined $\theta$] and take
    $\lambda_4-\lambda_1=1>0$, while $\lambda_1$ is increased.
    The Josephson coupling is decreased for increasing $\lambda_1$ in
    order to retain well-defined VEVs of the fields (vacuum expectation
    values); viz.~it is taken to be $\gamma_1=\frac{1}{4\lambda_1}$. 
    (d) The DW energy (tension) in the sigma-model limit (solid red line) is compared to that of the full numerical computations (points).
    In this figure $m=\sqrt{2}$, $\lambda_{2,3}=0$,
    $\lambda_4=\lambda_1+1$, $\gamma_1=\frac{1}{4\lambda_1}$ and
    $\gamma_{2,3}=0$.
  }
  \label{fig:g12vac}
\end{figure}
In fig.~\ref{fig:g12vac} the full numerical computations with
solid lines are compared with the numerical solution in the sigma-model limit,
which is a solution to eq.~\eqref{eq:eom_gamma1_theta_simpl}.
The ground state is only well defined for small enough
$\gamma_1$ and the square root in eq.~\eqref{eq:vac_gamma1_dw} is only
well defined in the $\lambda_1\to\infty$ limit if $\gamma_1\lambda_1$
is kept fixed.
Hence, we will set $\gamma_1=\frac{1}{4\lambda_1}$ which will keep the
asymmetric ground state well defined for the values of the parameters
chosen in fig.~\ref{fig:g12vac}.
Fig.~\ref{fig:g12vac}(a) displays the scalar fields but for larger
values of the coupling $\lambda_1$, the solid and dashed lines do not
really converge; hence the sigma-model limit does not provide accurate
approximations for large values of $\lambda_1$, when $\gamma_1$ is
decreased as $\gamma_1=\frac{1}{4\lambda_1}$.
In principle, the sigma-model limit would be a decent approximation for fixed $\gamma_1$, but such ground state does not exist in the large-$\lambda_1$ limit. 
On the other hand, fig.~\ref{fig:g12vac}(b) displays the sigma-model
constraint \eqref{eq:sigma_model_constraint}, which does indeed
converge to unity for large values of the coupling, $\lambda_1$.
We show the sigma-model limit solution, $\theta$, in fig.~\ref{fig:g12vac}(c) and compare the DW energies of the sigma-model limit and the full numerical computation in fig.~\ref{fig:g12vac}. 
As the sigma-model limit does not provide a good approximation for the limit with decreasing $\gamma_1$, also the energies are not well approximated by the sigma-model limit, although the overall trend of the large-$\lambda_1$ limit is captured by the sigma model approximation.

\subsection{Josephson chirally broken ground states
  \texorpdfstring{($\gamma_1\neq0$, $\gamma_2\in\mathbb{R}$, $\gamma_3=0$)}{(gamma1=/=0, gamma2 in R, gamma3=0)}: Kink at fine-tuned point}

The final example of numerical solutions is for the case of the
fine-tuned point in the theory, where the asymmetric ground state
becomes symmetric and their energies are degenerate.
This particularly special point in parameter space yields a very
special equation of motion in the sigma-model limit, which is
integrable and describes the kink soliton by the exact analytic
solution in the form of the arctan function.
We take $\gamma_1\neq0$, $\gamma_2=0$ for simplicity and $\gamma_3=0$
must be vanishing in this vacuum. 

\begin{figure}[!htp]
  \centering
  \mbox{\subfloat[]{\includegraphics[width=0.49\linewidth]{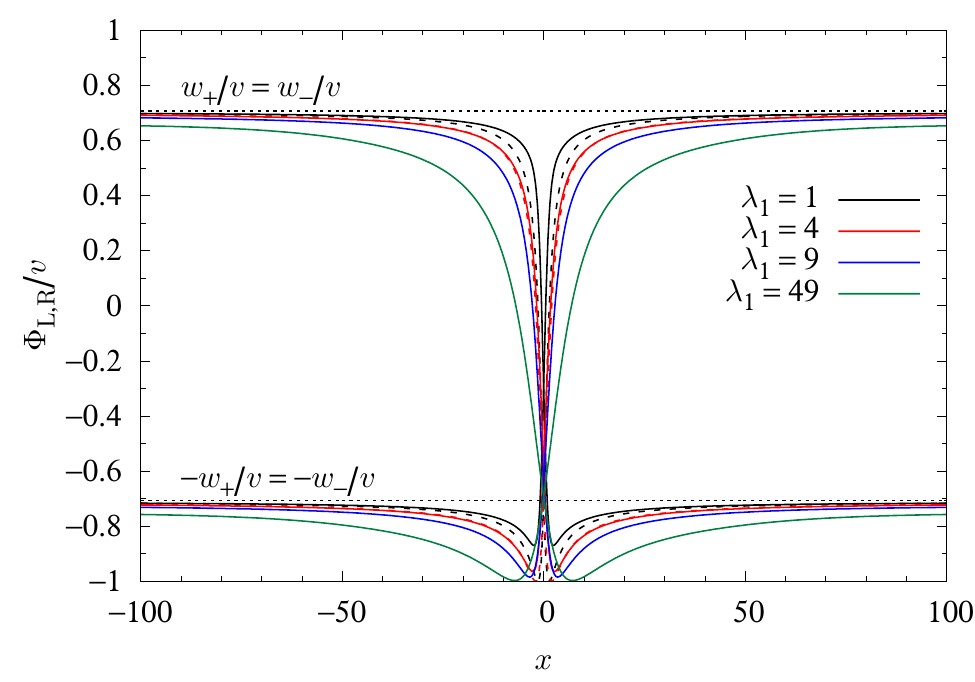}}
    \subfloat[]{\includegraphics[width=0.49\linewidth]{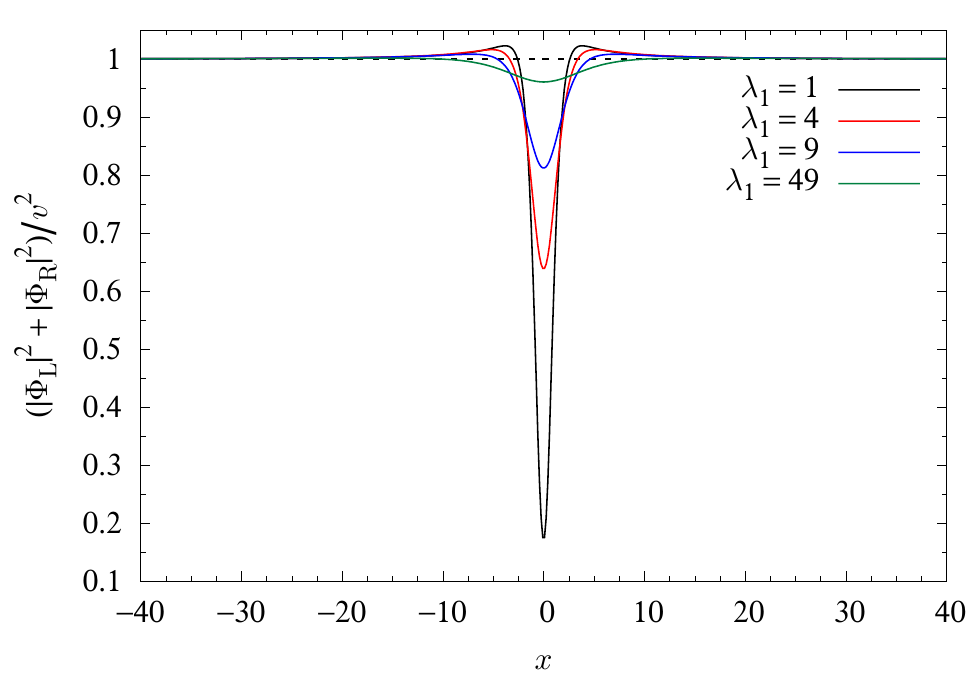}}}
  \mbox{\subfloat[]{\includegraphics[width=0.49\linewidth]{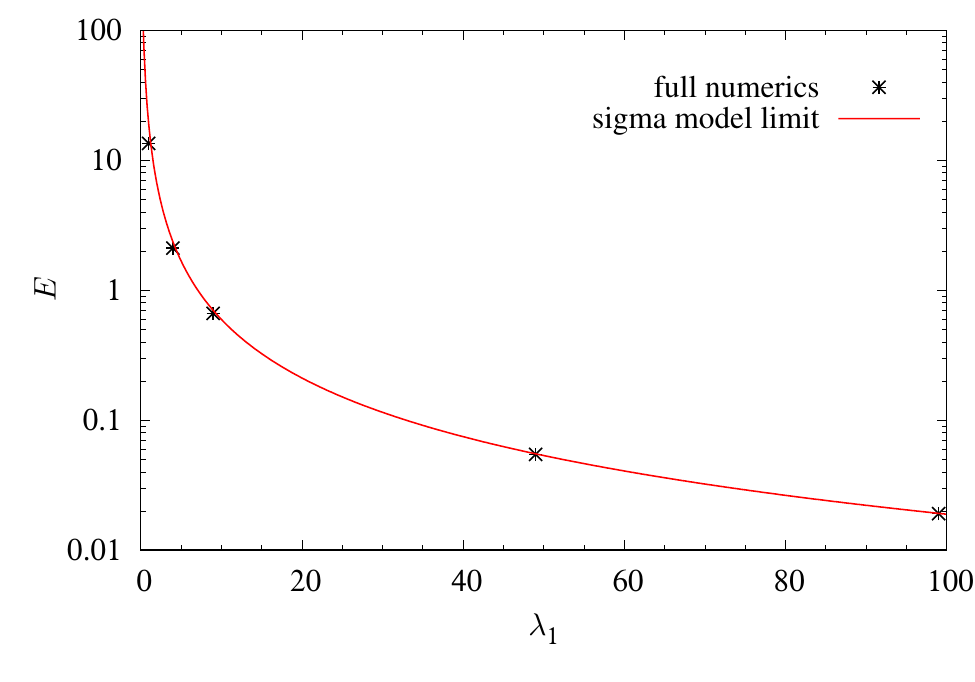}}}
  \caption{Kink solitons in the chirally broken fine-tuned Josephson 
    ground states interpolating between
    $(\Phi_\L,\Phi_\R)=\frac{1}{\sqrt{2}}(v\mathds{1}_3,-\sign(\gamma_1)v\mathds{1}_3)$ 
    and its sign-flipped sibling (see
    fig.~\ref{fig:Josephson_fine-tuned_vacua}), for a variety of
    couplings $\lambda_1=1,4,9,49,99$: (a) the diagonal part (all
    three elements are equal) of the scalar fields $\Phi_{\L,\R}$, and
    (b) the sigma-model constraint \eqref{eq:sigma_model_constraint}.
    We compare the full computations (solid lines) with
    the sigma-model limit (dashed lines) and take
    $\lambda_4-\lambda_1=1>0$, while $\lambda_1$ is increased.
    (c) Total energy (tension) of the DW with the solid line displaying the
sigma-model limit and the points showing the energies of full computations.
    The Josephson coupling is set to its critical value given by
    eq.~\eqref{eq:gamma1_critical}.
    In this figure $m=\sqrt{2}$, $\lambda_{2,3}=0$,
    $\lambda_4=\lambda_1+1$,
    $\gamma_1=\frac{m^2(\lambda_{34-1-2}+4\gamma_2)}{4\lambda_{12}}$
    and $\gamma_{2,3}=0$.}
  \label{fig:g12ftvac}
\end{figure}
In fig.~\ref{fig:g12ftvac}, the full numerical computations with
solid lines are compared to the exact analytical solution in the sigma-model limit \eqref{eq:kink_sol_fine_tuned} with dashed lines.
In the fine-tuned ground state, we have fixed the value of the
Josephson coupling $\gamma_1$ according to
eq.~\eqref{eq:gamma1_critical} with a positive sign.
Fig.~\ref{fig:g12ftvac}(a) displays the scalar fields and
fig.~\ref{fig:g12ftvac}(b) displays the sigma-model constraint
\eqref{eq:sigma_model_constraint}.
In this case, we see good convergence of the full numerical
computations to the sigma-model limit.
It is hence a good approximation for large $\lambda_1$ with
$\lambda_4-\lambda_1$ finite and $\gamma_1$ given by the critical
coupling (i.e.~eq.~\eqref{eq:gamma1_critical}).
Finally, we compute the DW energy (tension) of the sigma-model limit solution and the full numerical
computation and display them in fig.~\ref{fig:g12ftvac}(c) with a solid red line and points, respectively.
Since the profile functions converge well in fig.~\ref{fig:g12ftvac}(a) also the energy (tension) converges well in the large $\lambda_1$ limit (with $\lambda_4=\lambda_1+1$ fixed).

\section{Discussion and conclusion}\label{sec:conclusion}

In this paper, we have studied the unusual situation in which
the left- and the right-handed scalar field of the Ginzburg-Landau model attain different
vacuum-expectation values (VEVs) -- an asymmetric ground state.
We study four different scenarios, in both a sigma-model limit and
with full numerical computations of the equations of motion.
The sigma-model limit is given by sending both $\lambda_1$, the
coefficient of the single-trace ``$\phi^4$'' operator
$\Tr\big((\Phi_\L\Phi_\L^\dag)^2+(\Phi_\R\Phi_\R^\dag)\big)^2$, and
$\lambda_4$, the single-trace mixed-chirality operator
$\Tr(\Phi_\R\Phi_\L^\dag\Phi_\L\Phi_\R^\dag)$, to infinity
($\lambda_1\to\infty$ with $\lambda_4-\lambda_1$ finite and
positive).
The sigma-model limit has analytical and mathematical advantages, such
as allowing us in most cases to conclude that the scalar fields are
given by unitary matrices (times a constant) -- this is
rather helpful in establishing proofs and simplifications.
In many of the examples (except the Josephson kink soliton case and the case with nontrivial gauge fields) the
sigma-model limit turns out to be an excellent approximation for
$\lambda_1$ of the order of 50 or so.
In the case of the Josephson kink soliton, the ground state is only
well defined if the Josephson coupling is rather small, and not only:
it must obey the condition of $\gamma_1\lambda_1$ being fixed in the
sigma-model limit. This sends $\gamma_1$ to zero.
A peculiar property of this kink soliton is its near-fixed point of
its effective equation of motion, that moves closer to becoming a
fixed point as $\gamma_1$ tends to zero.
This prolongs the kink soliton structure and hence counteracts the
convergence of the true solution to the sigma-model limit
approximation.
The near-fixed point behavior of the equation of motion is
mathematically equivalent to what happens to the $\beta$-function in
walking technicolor theories \cite{Holdom:1984sk,Yamawaki:1985zg,Appelquist:1986an,Sannino:2004qp}.
In the $\beta$-function language, the conformal fixed point is the
vacuum in the sigma-model equation and the near-conformal fixed point
is exactly this non-existing vacuum (ground state) that prolongs the
kink soliton structure.
For the walking technicolor theories, this postpones the running of
the coupling for orders of magnitude in energy, thus keeping the
coupling at a semi-strong coupled value for a larger range in energy
scales.

Physically, there is no reason to believe that both the left- and the right-handed
quark condensates do not condense simultaneously and hence the domain-wall or
kink solitons studied in this paper are mere theoretical speculations.
Nevertheless, at very large densities it is not known exactly, from first principle-calculations, what
happens to the chiral Lagrangian Low-Energy Constants (LECs) and we leave the theoretical exotic phase of the Ginzburg-Landau theory as a nonperturbative option, that may or may not be relevant for QCD.

There are many avenues of continuing our research, many mathematical
proofs that could be made and finally many scenarios that we have not
even touched upon.
In particular, we have not determined the vacuum (ground state) for
the case with $\gamma_3\neq0$ and $\gamma_2>0$, as this entails
complex phases of at least one of the complex scalar fields, that are not even nice fractions of $\pi$.
We do not know the analytic solution to the vacuum equation in this
case.
There is also no reason to think that one of the three $\gamma$'s vanish.
However, we have not studied the case with all three $\gamma$ terms are
turned on, since it is the most complicated setting of the model and
we have not found analytic solutions to the vacuum equation in our
preliminary investigations, although with some conditions there may exist some.
A theoretical/mathematical problem is to study the sigma-model limit with the constraint $\Phi_\L^\dag\Phi_\L+\Phi_\R^\dag\Phi_\R=v^2\mathds{1}_3$ without assuming that both fields are unitary matrices multiplied by the constant (VEV) $v$.
In the DW vacuum where one of the scalar fields ($\Phi_\L$ or $\Phi_\R$) vanishes -- which is only the case without the Josephson ($\gamma_1$) term turned on -- this assumption is sound. But in the middle of the DW or in the case of the Josephson term turned on, this assumption may be relaxed.
Further development in this direction may improve the sigma-model limit of the theory exactly where it fails to be a good approximation.
We will, however, leave this direction for future studies.

Abelian Abrikosov-Nielsen-Olesen (ANO) vortices (or lump strings) can end on a DW in a $\U(1)$ gauge theory coupled to two complex scalar fields (or an $\Og(3)$ sigma model), and such a composite soliton is called a D-brane soliton \cite{Gauntlett:2000de,Shifman:2002jm,Isozumi:2004vg}. 
As a non-Abelian generalization, 
local non-Abelian vortices \cite{Hanany:2003hp,Auzzi:2003fs,Eto:2005yh} can end on non-Abelian domain walls \cite{Shifman:2003uh}.
Similarly, as a global $\U(1)$ counterpart, 
non-Abelian vortices 
\cite{Eto:2013hoa,Balachandran:2005ev,Nakano:2007dr,Eto:2009kg,Eto:2009bh,Eto:2009tr,Eto:2021nle,Gudnason:2025qxf} 
can end on 
the chiral non-Abelian domain walls explored in this paper, 
as a non-Abelian generalization of those in scalar two-component BECs
\cite{Kasamatsu:2010aq,Nitta:2012hy,Takeuchi:2012ee,Kasamatsu:2013lda,Kasamatsu:2013qia}. 
Exploring non-Abelian D-brane solitons in QCD or CFL interfaces remains as one of the challenging problems.

\subsection*{Acknowledgments}

S.~B.~G.~thanks the Outstanding Talent Program of Henan University for partial support.
The work of M.~N.~is supported in part by JSPS KAKENHI [Grants Nos.~JP22H01221 and JP23K22492], and the
WPI program ``Sustainability with Knotted Chiral Meta
Matter (WPI-SKCM$^2$)'' at Hiroshima University.

\bibliographystyle{JHEP}
\bibliography{bib_chiral}

\end{document}